\documentclass[pr,twocolumn,preprintnumbers,nofootinbib,showpacs,amsmath,amssymb,superscriptaddress]{revtex4-1}
\pdfoutput=1
\usepackage{psfrag,graphicx}
\usepackage[
      colorlinks=true,
      linkcolor=blue,
      urlcolor=blue,
      filecolor=black,
      citecolor=blue,
            ]{hyperref}
\usepackage{color}
\usepackage{amsfonts}
\usepackage{amssymb}
\usepackage{epstopdf}
\newcommand{\td}{\text{d}}
\newcommand{\el}{\ell_{\text{AdS}}}

\begin{document}
\title{Strong energy condition and complexity growth bound in holography}
\author{Run-Qiu Yang}
\email{aqiu@kias.re.kr}
\affiliation{Quantum Universe Center, Korea Institute for Advanced Study, Seoul 130-722, Korea}

\begin{abstract}
This paper proves that if eternal neutral black holes satisfy some general conditions and matter fields only appear in the outside of the Killing horizon, the strong energy condition is a sufficient condition to insure that the vacuum Schwarzschild black hole has the fastest action growth of the same total energy. This result is consistent with the bound of computational complexity growth rate and gives a strong evidence for the holographic complexity-action conjecture.
\end{abstract}
\maketitle

\noindent

\section{Introduction}
\begin{figure}
  \centering
  \includegraphics[width=0.27\textwidth]{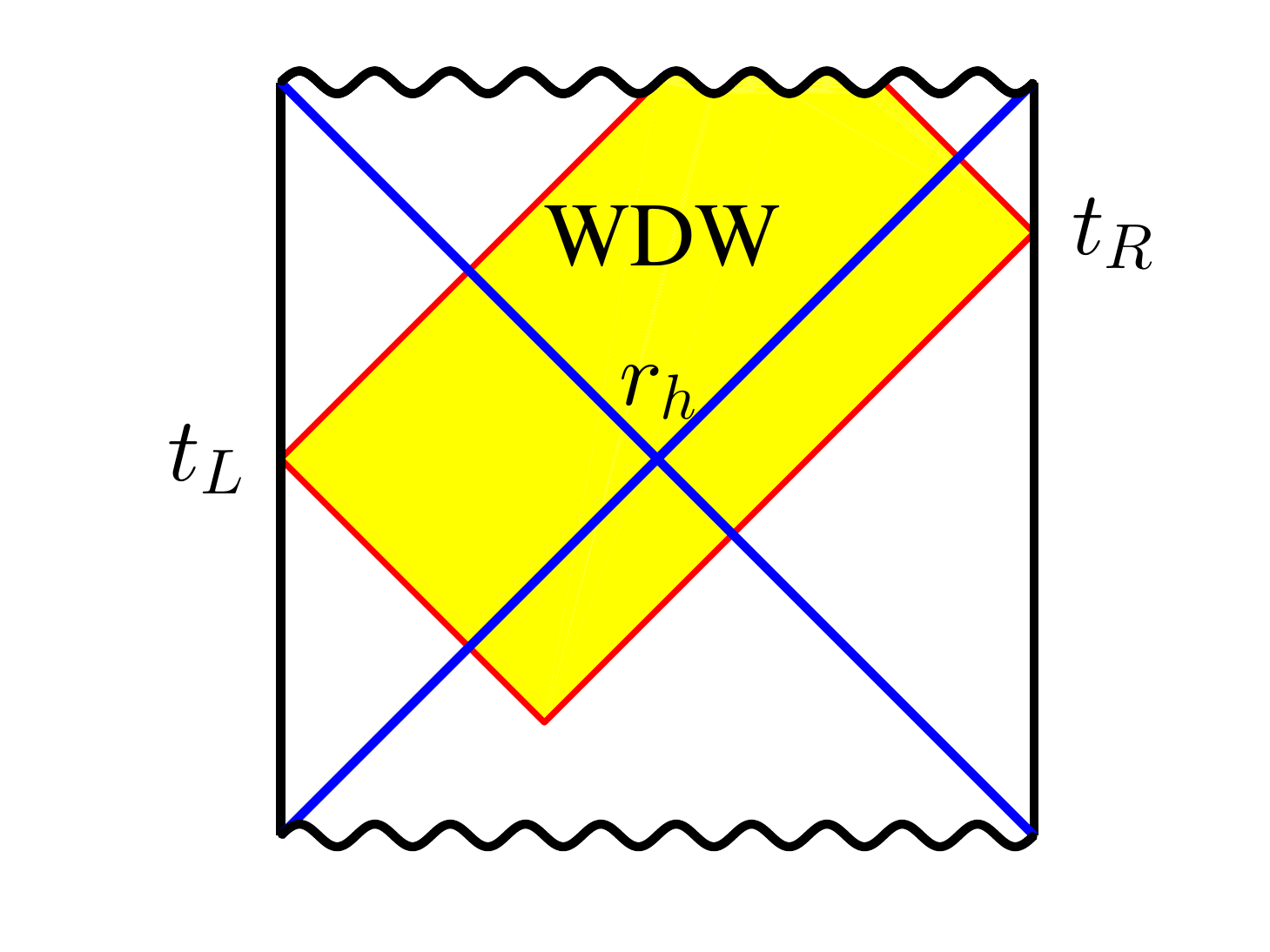}
  \caption{The Penrose diagram for Schwarzschild AdS black hole and the WDW patch. For given particular time slices at the two boundaries $t_L$ and $t_R$, the WDW patch is the yellow region with its boundary, which is just the domain of development of any spacelike slice which connects $t_L$ and $t_R$. }\label{SAdS}
\end{figure}

One exciting progress in quantum information theory and gravity theory is the discovery of a connection between computational complexity and inner dynamic of an asymptotic anti de-Sitter (AdS)  black hole. The authors in Refs. \cite{Brown:2015bva,Brown:2015lvg} proposed a complexity-action (CA) conjecture that the quantum complexity of a holographic state is dual to the inner dynamic of a black hole. More detailed, such conjecture can be is illustrated by the Fig.~\ref{SAdS}, where time slices $t_L$ and $t_R$ on the AdS boundaries of an eternal AdS-Schwarzschild black hole determine a particular quantum state~\cite{Maldacena:2001kr,Brown:2015lvg},
\begin{equation}\label{TFD1}
  |\psi(t_L,t_R)\rangle=e^{-i(H_Rt_R+H_Lt_L)}|\text{TFD}\rangle\,.
\end{equation}
Here $|\text{TFD}\rangle=Z^{-1/2}\sum_\alpha e^{-\beta E_\alpha/2}|E_\alpha\rangle|E_\alpha\rangle$ is the thermofield double state.\footnote{Note that with conventions in this paper, time on both sides of Fig.~\ref{SAdS}
increases with the same direction.}  CA conjecture states that the complexity of such particular state $|\psi(t_L,t_R)\rangle$ is given by the on-shell action in the Wheeler-DeWitt (WDW) patch,
\begin{equation}\label{CA-conj}
  \mathcal{C}(|\psi(t_L,t_R)\rangle):=\frac{\mathcal{A}}{\pi\hbar}\,.
\end{equation}
Here $\mathcal{A}$ is the on-shell action of dual gravity theory (with corresponding matter fields if dual boundary isn't vacuum) in WDW patch. The WDW patch is the domain of development of
any spacelike slice which intersects with the left boundary and right boundary at $t_L$ and $t_R$, e.g, see the Fig.~\ref{SAdS} as an example.  As the boundary of WDW patch has some null fragments, the CA conjecture faced obstacles on computing null boundary terms and the joints between null boundary and other boundaries when this conjecture was originally proposed. However, this problem has been overcome recently by carefully analyzing the variation problem involving null boundaries. For more detailed, one can see Refs.~\cite{Lehner:2016vdi}.

Although it may not be the main motivation for the authors in Refs.~\cite{Brown:2015bva,Brown:2015lvg}, one beautiful result and strong evidence to support CA conjecture is that it shows a new world record for black hole: the fastest computer. Complexity has two facets, information storage and information processing. For the former one facet, Bekenstein \cite{PhysRevD.7.2333} has shown that black holes set a theoretical maximum on information storage and no object stores more bits of information than a black hole of the same size. For the latter one, in the theory about computational complexity, there is an upper bound for a system containing mass $M$ (with light speed $c=1$) proposed by Refs.~\cite{MARGOLUS1998188,Lloyd2000},
\begin{equation}\label{growthbound}
  \frac{\td\mathcal{C}}{\td t}\leq\frac{2M}{\pi\hbar}.
\end{equation}
The authors in Refs. \cite{Brown:2015bva,Brown:2015lvg} checked it according to CA conjecture by some black holes, and found that it was saturated exactly for some black holes  at the late time limit (i.e., the limit of $t_L\rightarrow\infty$ or $t_R\rightarrow\infty$).
In this paper, the attention is concentrated on the bound equation \eqref{growthbound} itself. By the combination of CA conjecture and complexity growth rate bound, one may expect a new universal bound in the gravity theory,
\begin{equation}\label{growthbound2}
  \frac{\td\mathcal{A}}{\td t}\leq2M
\end{equation}
at least for stationary asymptotic AdS space-time at the late time limit.
The inequality \eqref{growthbound2} was born from the considerations about quantum gravity and AdS/CFT correspondence, but it only contains some quantities in the classical gravity theory. As what we have seen that, though the thermodynamical properties of black hole are the results of quantum aspects of gravity, they can obtain very strict proofs in classical general relativity. One may naturally ask that if such thing can happen for the bound equation \eqref{growthbound2}. As both sides of inequality \eqref{growthbound2} only involve gravity quantities, it should be proven from the gravity theory itself if the CA conjecture is correct. However, there isn't any strict discussion about it yet.

Though Ref.~\cite{Brown:2015lvg} gave the first example to check the inequality \eqref{growthbound2}, where a static shell was added in the outer region of horizon, it is too special to stand for general cases. For an eternal black hole with other general conditions, this paper will show that, if matter fields only appear in the outside of the Killing horizon and satisfy the strong energy condition, the inequality \eqref{growthbound2} is always true at the late time limit and the vacuum Schwarzschild black hole can saturate inequality \eqref{growthbound2}.

\section{Conditions for matter fields}\label{spher}
Before we discuss what restrictions should be added on the matter fields so that the inequality \eqref{growthbound2} can always be true, let's first take some time to consider the inequality \eqref{growthbound2} itself. For any two time slices at the left and right boundaries respectively, the on-shell action in the corresponding WDW patch is coordinately independent, but its growth rate, the derivative with respective to time, is coordinately dependent. For an eternal black hole, as it has a Killing vector $\xi^\mu$ which is timelike at the boundary, one physical feasible method to define the left-hand of  the inequality \eqref{growthbound2} is to demand that $t$ is the integral parameter of $\xi^\mu$, i.e., $(\partial/\partial t)^\mu=\xi^\mu$. Such definition about the time derivative sets the left-hand of the inequality \eqref{growthbound2} involving the choice of Killing vector $\xi^\mu$. On the other hand, the right-hand of  the inequality \eqref{growthbound2} involves the the mass of a black hole, which is also coordinately dependent and has several different definitions in the asymptotic AdS space-time such as ADM mass \cite{Abbott:1981ff}, Misner-Sharp mass \cite{PhysRev.136.B571}, Komar mass \cite{PhysRev.129.1873}, quasilocal energy \cite{Brown:1992br} and so on. Even in the static spherical case, if some matter fields appear, different mass definitions may give different results.
In the isolated asymptotic AdS black hole, which means that the energy momentum tensor field ${T^\mu}_\nu$ satisfies $|r^3{T^\mu}_\nu|<\infty$, the metric can always have the following form,
\begin{equation}\label{metric0}
  \td s^2=-f(r)\td t^2+f(r)^{-1}\td r^2+r^2\td\Omega^2_k
\end{equation}
with $f(r)=k+r^2/\el^2-2f_0/r+\mathcal{O}(1/r^2)$. Here $r$ is the radius coordinate, $\el$ is the AdS radius, $k=0,\pm1$ and $\td\Omega^2_k$ is the line element in the dual boundary. Thus, all the physical acceptable definitions about mass (or mass density) should give $M=f_0$. One candidate to compute the mass in the inequality \eqref{growthbound2} in this case is the Komar mass. For any three dimensional hypersurface $\Sigma$, which can be spacelike, timelike or mixed with these two cases, with its boundary $S$, one can define the Komar mass on it as follws \cite{Barnich:2004uw,Kastor:2009wy},
\begin{equation}\label{Smarr2}
\begin{split}
   m&=-\frac1{8\pi}\oint_{S}\left(\nabla^\mu\xi^\nu-\frac3{\el^2}\omega^{\mu\nu}\right)\td S_{\mu\nu}\\
   &=2\int_{\Sigma}(T_{\mu\nu}-\frac12Tg_{\mu\nu})n^\mu\xi^\nu\td V\,.
   \end{split}
\end{equation}
Here $\td S_{\mu\nu}$ is the directed surface element and $\omega^{\mu\nu}$ is the Killing potential satisfied $\xi^\mu=\nabla_\nu\omega^{\nu\mu}$. $n^\mu$ is the unit normal vector field for $\Sigma$ and $\td V$ is volume element of $\Sigma$. $n^\mu$ is future directed if it is timelike and inward directed if it is spacelike. The Killing potential isn't unique, but different choices can give the same integration result \cite{Kastor:2009wy}. The total energy for the system is $M=m(S\rightarrow\infty)$.\footnote{If the consideration is restricted in the framework of AdS/CFT correspondence, then there is a special notion of time and energy that can be computed in terms of the boundary Hamiltonian with respect to this time. For the static asymptotic isolated case, this special time is just the integral parameter of $\xi^\mu$ and the energy is as the same as $m(S\rightarrow\infty)$. However, in this paper, $\xi^\mu$ is a general Killing vector field which is timelike at the boundary.}

Let's begin the discussion from very simple case, where the space-time has spherical symmetry and matter fields only distribute outside of the horizon. The metric for spherical static space-time can be written as the following form,
\begin{equation}\label{metric1}
  \td s^2=-f(r)e^{-\chi(r)}\td t^2+f(r)^{-1}\td r^2+r^2\td\Omega^2.
\end{equation}
%
Here $f(r)$ is positive when $r>r_h$ and negative when $r<r_h$. A horizon locates at $r=r_h$.

To warm up, let's first review the discussions shown by Adam R. Brown et al. in Refs. \cite{Brown:2015lvg,Brown:2015bva} by adding a static shell at $r=r_0$, which is considered as a very important evidence for CA conjecture. Suppose that the shell carries some mass so that we have the following solution for metric,
\begin{equation}\label{solu0}
\begin{split}
f(r)&=1+r^2/\el-2[M_0+\delta M\Theta(r-r_0)]/r,\\
\chi(r)&=\tilde{\chi}-\Theta(r-r_0){\tilde{\chi}}\,.
\end{split}
\end{equation}
with $\tilde{\chi}=-\left[\ln f(r)\right]_{r_0}$. Here the notation $[X]_{r_0}$ stands for $X(r_0^+)-X(r_0^-)$ and $\Theta(r-r_0)$ is the step function.
Then the total mass is $M=M_0+\delta M$. Based on the discussions in Ref.~\cite{Brown:2015bva}, one can find the action growth rate is,
\begin{equation}\label{grwothrate1}
  \frac{\td \mathcal{A}}{\td t}=2e^{-\tilde{\chi}/2}M_0=2m_H.
\end{equation}
Here $m_H$ is the Komar mass at the horizon. Brown et al. argued that $\frac{\td \mathcal{A}}{\td t}<2M$ by implying an assumption $\delta M>0$ \cite{Brown:2015lvg,Brown:2015bva}. Though this assumption is very physically believable, it is still a restriction on matters. Once we no longer impose this assumption, for example, by setting $\delta M<0$, then we find that inequality \eqref{growthbound2} is violated.

In fact, even keeping the energy density of the matter positive, the violation can still appear. To see this, let's consider the case that there is a nonzero energy momentum tensor field $T^{\mu\nu}$ at the region of $\tilde{r}>r_0$, and it is found that,
\begin{equation}\label{solu2}
   m(\tilde{r})=m_H+2\int_{\Sigma(r_0<r<\tilde{r})}(T_{\mu\nu}-\frac12Tg_{\mu\nu})n^\mu\xi^\nu\td V\,.
\end{equation}
with $n^\mu=\xi^\mu/\sqrt{-\xi^\alpha\xi_\alpha}$.
According to a similar argument about Eq.~\eqref{grwothrate1}, we can find that the action growth rate is expressed as the same as Eq.~\eqref{grwothrate1}.
In some region of $\Sigma$ where the pressure is too negative so that $(T_{\mu\nu}-\frac12Tg_{\mu\nu})\xi^\mu\xi^\nu<0$, then one may obtain $M=m(\infty)<m_H$. In this case, one can find that the inequality \eqref{growthbound2} is broken. In order to insure the inequality \eqref{growthbound2}, we need that $M\geq m_H$. This leads to,
\begin{equation}\label{strongcd1}
  \int_{\Sigma(r>r_0)}(T_{\mu\nu}-\frac12Tg_{\mu\nu})\xi^\mu\xi^\nu\td V\geq0
\end{equation}
for timelike Killing vector field $\xi^\mu$. One sufficient condition to insure the inequality \eqref{strongcd1} is ``strong energy condition". The strong energy condition is the sufficient condition for the inequality \eqref{strongcd1} because of two points: (1) the strong energy condition is a local energy condition and leads to $(T_{\mu\nu}-\frac12Tg_{\mu\nu})\xi^\mu\xi^\nu\geq0$ holes at every point in the outside of the horizon but the inequality \eqref{strongcd1} is about the integration of $(T_{\mu\nu}-\frac12Tg_{\mu\nu})\xi^\mu\xi^\nu$ and (2) the strong energy condition will lead the inequality \eqref{strongcd1} holding for any time like vector field $\xi^\mu$ rather than only for timelike Killing vector fields. It is easy to see that the saturation appears when $(T_{\mu\nu}-\frac12Tg_{\mu\nu})\xi^\mu\xi^\nu=0$ for  the timelike Killing vector field $\xi^\mu$. This means that the vacuum Schwarzschild black hole is one which can saturate the inequality \eqref{growthbound2} for the given mass $M$ in spherical static systems.



\section{General cases}\label{gencase}
After giving a sufficient condition to insure the inequality \eqref{growthbound2} in the spherical case, let's consider general static asymptotic AdS$_4$ black holes, where the matter fields still distribute in the outside of the outmost non-degenerated Killing horizon (the definition of ``outmost non-degenerated Killing horizon'' will be clarified later on), but the space-time is a lack of symmetry. Now as the geometry behind the horizon may be very arbitrary, the causal property and topology of singularity are free. This leads to the fact that the definition of the WDW-patch and calculation about action in it are ambiguous as they strongly depend on the casual structure of space-time.
In order to prove the inequality \eqref{growthbound2}, let's first discuss the properties of the Killing horizon and singularities of space-time (after maximum extension) and then introduce some assumptions.

Let $\mathcal{M}$  stand for a static maximal extension asymptotic AdS space-time with a Killing vector field $\xi^\mu=(\partial/\partial t)^\mu$ which is timelike at the AdS boundaries $B_L$ and $B_R$. The boundaries $B_L$ and $B_R$ are not in the real space-time, but here they are added into $\mathcal{M}$ for convenience. As the space-time is static, a set of hypersurfaces is the Killing horizon corresponding to $\xi^\mu$ if and only if $\xi^\mu\xi_\mu$ is zero on it. In following, when we say ``Killing horizon,'' it always means the Killing horizon corresponding to $\xi^\mu=(\partial/\partial t)^\mu$.  Take $\mathcal{K} =\{H_1,H_2,\cdots\}$, where $H_i$ is one nondegenerated connected branch of the Killing horizon. Here ``nondegenerated" means the square of Killing vector $\xi^2$ has different signs in the two sides of $H_i$.  As the space-time is static, $H_i$ is a bifurcated Killing horizon and does not contain future and pass endpoints.  For a space-time, there may be no Killing horizon so $\mathcal{K} =\emptyset$. Even in the cases that $\mathcal{M}$ contains the Killing horizon, $\mathcal{K} $ can still be empty set if every branch of the Killing horizon is degenerated. A branch in $\mathcal{K} $ is called outmost nondegenerated Killing horizon (ONKH) if and only if all future and past null geodesics coming from $B_L$ and $B_R$ can reach it without touching any other branches of $\mathcal{K} $. And a branch in $\mathcal{K} $ is called sub-outmost nondegenerated Killing horizon (sub-ONKH) if and only if all future and past null geodesics coming from $B_L$ and $B_R$ can reach it and except for the ONKH they do not touch any other branches of $\mathcal{K} $.

For a general space-time, the $\mathcal{K} $ may be nonempty but it does not contain the ONKH or sub-ONKH. For example, if we add some matter fields in the outside of the Killing horizon for Schwarzschild AdS black hole, then a new singularity may appear without adding new Killing horizon or causing any effect on the original Killing horizon. Then some null geodesics coming from the boundary cannot reach the Killing horizon. For the cases that the Killing horizon has very complicated geometrical structure, it is possible that some null geodesics will first meet a branch but some others will first meet the other branches.  Assume $H_o$ and $H_{\text{sub-o}}$ stand for the ONKH and sub-ONKH. Now let's introduce an assumption on $\mathcal{K} $:

\textit{if $\mathcal{K} \neq\emptyset$, then $H_o\in \mathcal{K}$ and $\{H_o\}\subsetneq \mathcal{K}$ only when $H_{\text{sub-o}}\ \in \mathcal{K}$.} \\
This assumption shows that if there is a nondegenerated Killing horizon then all the singularities hide behind the ONKH or sub-ONKH. It also means that the causal structures in the outside of ONKH is trivial, i.e., the space-time in this region is regular and $\xi^\mu$ is always timelike except for some regular zero measurement surfaces (the degenerated Killing horizons).

In the following, it will be shown that the inequality \eqref{growthbound2} is always true in a general static black hole when (i) matter fields appear in the outside of ONKH and satisfy strong energy condition, and (ii) the black hole satisfies one of following four conditions:\footnote{There is a possibility that $\mathcal{K} =\{H_o\}$ and space-time has no singularity. We don't consider this case as such space-time is not a black hole.}\\
 \textit{ (a) $\mathcal{K} =\emptyset$ and the total mass $M\geq0$;\\
 (b1) $\mathcal{K} =\{H_o\}$ and $\xi^\mu$ becomes null in the singularities (in the sense of limit) and $M\geq0$; \\
 (b2) $\mathcal{K} =\{H_o\}$ and the union of singularities has $R\times S^2$ topology and $\xi^\mu\xi_\mu\neq0$ in it (in the sense of limit); \\
 (c) $H_{\text{sub-o}}\in \mathcal{K}$ and total mass $M\geq0$.} \\

Before we give the proofs, it is worth explaining a few things. A static space-time which satisfies the condition on $\mathcal{K} $ has only three cases: $\mathcal{K} =\emptyset$, $\mathcal{K} =\{H_o\}$ and $H_{\text{sub-o}}\in \mathcal{K}$. For the static space-time, there are some black holes which may have apparent horizon, Killing horizon or event horizon but have no any nondegenerated Killing horizon. Such cases can satisfy the case (a). It will be shown later that the condition $\xi^2=0$ at the singularity implies that the singularity is null in the case (b1). The case (b2) does not need the total mass $M$ is non-negative. In the case (b2), though the topology of the singularities is $R\times S^2$, the topology of the whole space-time may not be $R\times S^2$. This is because that the space-time in the outside of ONKH is static and only the inner region of ONKH contributes on the action growth rate. In the statement of case (b2), it does not explicitly demand the singularities are spacelike, timelike or null, however, we will see that the case (b2) leads that the singularities can only be spacelike. $\xi^2$ in the statement (b2) is nonzero, because if $\xi^2$ is zero then this case is belong to case (b1). In the case (b1) and (c), there is not any special demand on the singularity. We will see later that this is because the WDW patch will not touch the singularities.

In the case (a), as the whole space-time is static except for some zero measurement null hypersurface, action growth rate is zero and bound inequality~\eqref{growthbound2} is correct. As the most complicated case is (b2), Let's first give the proof for case (b2) and then give the proofs for cases (b1) and (c).

\subsection{Proof for case (b2)}\label{spacebd}
As the same as previous discussions, the action growths in the outside of ONKH are canceled with each others. Let $\mathcal{A}_{\text{in,bulk}}$, $\mathcal{A}_{\text{in,bd}}$ and $\mathcal{A}_{\text{in,joint}}$ stands for the three different kinds of contributions in the region behind the ONKH, i.e.,  the contributions from the bulk integration, the boundary term and the joint term respectively. Then we can find,
\begin{equation}\label{dsdt1}
 \frac{\td \mathcal{A}}{\td t}= \frac{\td }{\td t}\mathcal{A}_{\text{in,bulk}}+ \frac{\td }{\td t}\mathcal{A}_{\text{in,bd}}+ \frac{\td }{\td t}\mathcal{A}_{\text{in,joint}}\,.
 \end{equation}
The suitable boundary term and joint term have been discussed and given in Refs. \cite{Parattu:2015gga,Lehner:2016vdi}.

\subsubsection{Bulk contribution}\label{b2bulk}
Let's first consider the bulk term. By using Einstein's equation, we can prove that the action growth coming from the bulk integration can be converted into following integration (see appendix~\ref{app1}),
\begin{equation}\label{bulkL}
  \frac{\td }{\td t}\mathcal{A}_{\text{in,bulk}}=-\frac3{8\pi\el^2}\int_{\Sigma'} \sqrt{\xi^\mu\xi_\mu}\td V\,.
\end{equation}
Here $\Sigma'$ stands for the timelike 3-hypersurface in inner region of ONKH, which is orthogonal to $\xi^\mu$ and connects the ONKH and singularity (see Fig.~\eqref{Fig2}).

\begin{figure}
  \centering
  \includegraphics[width=0.25\textwidth]{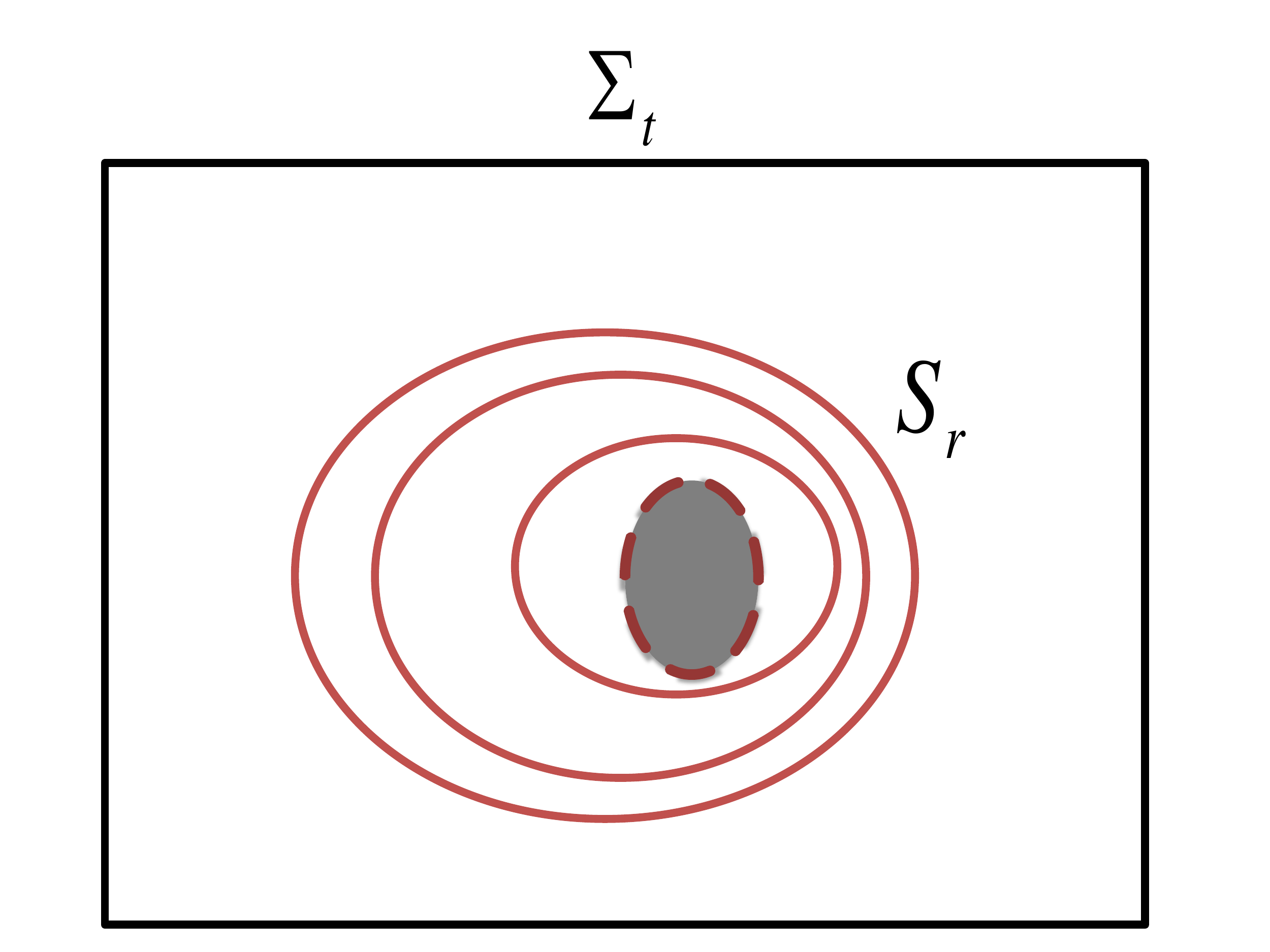}
  \caption{The 2+1 decomposition of $\Sigma_t$ near the singularity. The red circles stands for the surfaces $\Psi=r$ and the red dashed circle stand for the singulary $\Psi=0$.   }\label{Fig1}
\end{figure}

\subsubsection{Contribution of singularities}\label{b2s}
Based on the results in Ref.~\cite{Lehner:2016vdi}, the only boundary term comes from the singularity.\footnote{The null boundary term used in Ref.~\cite{Lehner:2016vdi} depends on the choices of parametrization on bull boundary.  Ref.~\cite{Lehner:2016vdi} also advises adding an additional null boundary term to eliminate this dependence. However, this additional null boundary term has no contribution on the action at the late time limit.} Now let's try to find such contribution. As Killing vector $\xi^\mu$ is  hyperorthogonal, near the singularity we can always find a coordinate $\{t,\vec{x}\}$ such that the metric has following form,
\begin{equation}\label{decomp1}
  \td s^2=N^2(\vec{x})\td t^2+\td s_{\Sigma_t}^2\,.
\end{equation}
where $\Sigma_t$ is the timelike 3-surface which is orthogonal to $\xi^\mu$. The unit normal of $\Sigma_t$ then is $n^\mu=\xi^\mu/N$ with $N=\sqrt{\xi^\mu\xi_\mu}$. $\td s_{\Sigma_t}^2$ is the induced line element on $\Sigma_t$, which is independent of the coordinate $t$. 
The collection of all the singularities in a maximal extension space-time  are some 3 dimensional hypersurfaces in general. We can find a function $\Psi(t,\vec{x})=0$ to describe the positions of these singularities. As the metric components are independent of $t$, we find $\Psi(t,\vec{x})=\Psi(\vec{x})$,
which gives two-dimensional surfaces in the submanifold $\Sigma_t$. As the topology of singularities is $R\times S^2$, these singular 2-surfaces must be connected and topology homomorphism to a unit sphere, otherwise, the topology of singularities isn't $R\times S^2$.


Let's first assume the singularities are non-null. It will shown later that null singularity cannot appear in the case (b2). As shown in the Fig.~\ref{Fig1}, $S_0$ is the singular 2 dimensional surface, which is spacelike or timelike. Then we decompose $\Sigma_t$ by a series closed 2-surfaces $\{S_r\}$ with $r\geq0$ and every 2-surface $S_r$ is given by a function $\Psi(\vec{x})=r$. One can build a local coordinate on every closed 2-surface by spherical coordinate $\{\theta, \phi\}$. This gives a 2+1 decomposition on $\Sigma_t$ by coordinate $\{r, \theta,\phi\}$ and the line element can be written as,
\begin{equation}\label{decomp2}
  \td s_{\Sigma_t}^2=-\varepsilon F^2\td r^2+h_{AB}(\td x^A+l^A\td r)(\td x^B+l^B\td r)
\end{equation}
with $A, B=$ 1 or 2 and $\varepsilon\equiv\pm1$. When the singularity is spacelike, $\varepsilon=1$, otherwise $\varepsilon=-1$. $F, h_{AB}$ and $l^A$ are all the function of $\vec{x}$. The singularity happens at $r=0$.
By the gauge freedom, we can choose that $l^A=0$ and $F=1$. This gauge can be obtained by use Gaussian normal coordinate system. Now in the coordinate $\{t,r,\theta,\phi\}$, the line element near the singularity can be written as,
\begin{equation}\label{decomp3}
  \td s^2=-\varepsilon\td r^2+N^2\td t^2+h_{AB}\td x^A\td x^B.
\end{equation}

At the boundary $r=\epsilon\rightarrow0$, as it is non-null, the boundary term is just the Gibbosn-Hawking-York (GHK)term$-\frac{\varepsilon}{8\pi}\int K dV$, where $K$ is the trace of extrinsic curvature of the non-null hypersurface. In order to compute the such contribution, one can use 3-surface $r=\epsilon$ to warp up the singularity and compute the GHK term at this boundary. After some computations, one can find that,
\begin{equation}\label{bdS0a}
  \frac{\td }{\td t}\mathcal{A}_{\text{in,bd}}(r)=\frac{\varepsilon}{8\pi}\oint \partial_r(N\sqrt{|h|})\td^2x.
\end{equation}
We can compare it with Komar mass contained in the surface $r=\epsilon$, which is,
\begin{equation}\label{bdS0b}
  m=\frac{-\varepsilon}{8\pi}\oint \frac{\partial_rN^2}{N}\sqrt{|h|}\td^2x-\frac3{4\pi\el^2}\int_{\Sigma_{r<\epsilon}} N\sqrt{|h|}\td r\td^2x.
\end{equation}
Here $\Sigma_{(r<\epsilon)}$ stands for  the inner region bounded by $r=\epsilon$ surface in $\Sigma_t$. The details about Eqs.~\eqref{bdS0a} and \eqref{bdS0b} can be found in appendix~\ref{app2}. As the matter fields only appear the outside of $H_o$, the formula \eqref{Smarr2} shows the Komar mass contained in the singularity equals to the mass contained in $H_o$, i.e., $m(r\rightarrow0)=m_H$. By some algebras, we can get,
\begin{equation}\label{bdS0c}
\begin{split}
  &\frac{\td }{\td t}\mathcal{A}_{\text{in,bd}}(r)-\frac32m_H-\frac9{8\pi\el^2}\int_{\Sigma_{r<\epsilon}} N\sqrt{|h|}\td r\td^2x\\
  =&\frac1{8\pi}\oint[2\varepsilon\sqrt{|h|}\partial_rN^2+\varepsilon N^2\partial_r\sqrt{|h|}]N^{-1}\td^2x\,.
  \end{split}
\end{equation}
Now we need to know the asymptotic behaviors of function $N$ and metric $h_{AB}$ when $r\rightarrow0$.

When $r\rightarrow0$, the leading terms of $N$ and metric $h_{AB}$ can be written as $N=r^{p_t}\sqrt{f_1(\theta,\phi)}$ and $h_{AB}=r^{2p_{AB}}\tilde{h}_{AB}(\theta,\phi)$ (no summation) with smooth functions $f_1$ and $\tilde{h}_{AB}$.\footnote{Here it is assumed that $p_t$ and $p_{AB}$ are independent of $\theta$ and $\phi$ but may be the functions of $r$.} As the surface $r=$constant in $\Sigma_t$ is topological homeomorphism to a unit sphere, its scalar curvature $\mathcal{R}^{(2)}$ satisfies that $\oint {\mathcal{R}^{(2)}}\sqrt{|h|}\td^2x=8\pi$ for any nonzero $r$. One can check that this leads that $p_{AB}=p_0$ and leading term of the line element \eqref{decomp3} can be written as,
\begin{equation}\label{decomp4}
  \td s^2=-\varepsilon\td r^2+r^{2p_t}f_1(\theta,\phi)\td t^2+r^{2p_0}f_2(\theta,\phi)\td\Omega^2
\end{equation}
for some suitable coordinate $\{\theta, \phi\}$ and smooth function $f_2(\theta,\phi)$.  As the line element is Lorentzian, the sign $\varepsilon=1$. This means that singularities cannot be timelike. Now take this metric into Einstein's equation, one can find that the system can self-consistent only when $f_1(\theta,\phi)=$constant and $p_0=2/3, p_t=-1/3$. Then we obtain that,
\begin{equation}\label{bdS1}
  \frac{\td }{\td t}\mathcal{A}_{\text{in,bd}}=\frac32m_H.
\end{equation}
For details, one can see appendix \ref{app3}. One interesting thing is that the metric \eqref{decomp4} is similar to the form of generical Kasner's metric \cite{Belinskii1970,Belinskii1982} and the indexs $p_0$ and $p_t$ satisfy the Kasner's relationships $2p_0+p_t=2p_0^2+p_t^2=1$.

Now let's show that the null singularity cannot appear in the case (b2). Let $S_0$ stand for the union of all singular 3-dimensional surfaces and take $\psi_t$ to be the tangent map generated by Killing vector $\xi^\mu$. As $\xi^\mu$ is a Killing vector field, $\psi_t[S_0]$ must be also some null singular 3-dimensional surfaces and $\psi_t[S_0]\subseteq S_0$. This means that $\psi_t$ is a differential homeomorphic map from $S_0$ into itself, so $\xi^\mu$ is tangent to $S_0$. In this case, the coordinate line $t$ lays on the null singular surface $S_0$. Assume $\{t,x^1,x^2\}$ to be a local coordinate on $S_0$, which dose not need to cover the whole region of $S_0$. As there is time reversal symmetry, the induced metric on $S_0$ has following form,
\begin{equation}\label{dsnull1}
  \td s^2_{S_0}=N^2\td t^2+\sigma_{AB}\td x^A\td x^B\,.
\end{equation}
Here $\sigma$ is only the function of $\{x^1,x^2\}$ and $N\neq0$ according to the condition in case (b2). As $S_0$ is null, the determinant of this metric is zero, i.e., $\text{det}[\sigma_{AB}]=0$. Let $\Sigma_t$ be the hypersurface of fixed time $t$ in $S_0$, then the induced metric is,
\begin{equation}\label{dsnull2}
  \td s^2_{\Xi_t}=\sigma_{AB}\td x^A\td x^B\,.
\end{equation}
As the topology of $S_0$ is $R\times S^2$,  $\Xi_t$ must have topology $S^2$. However, any manifold which has topology of $S^2$ must be conformal to unit sphere locally, so $\Xi_t$ is spacelike and det$[\sigma_{AB}]>0$. This is self-contradiction and so the singular surfaces cannot be null in the case (b2).

\subsubsection{Contributions of joints}\label{b2joint}
\begin{figure}
  \centering
  \includegraphics[width=0.33\textwidth]{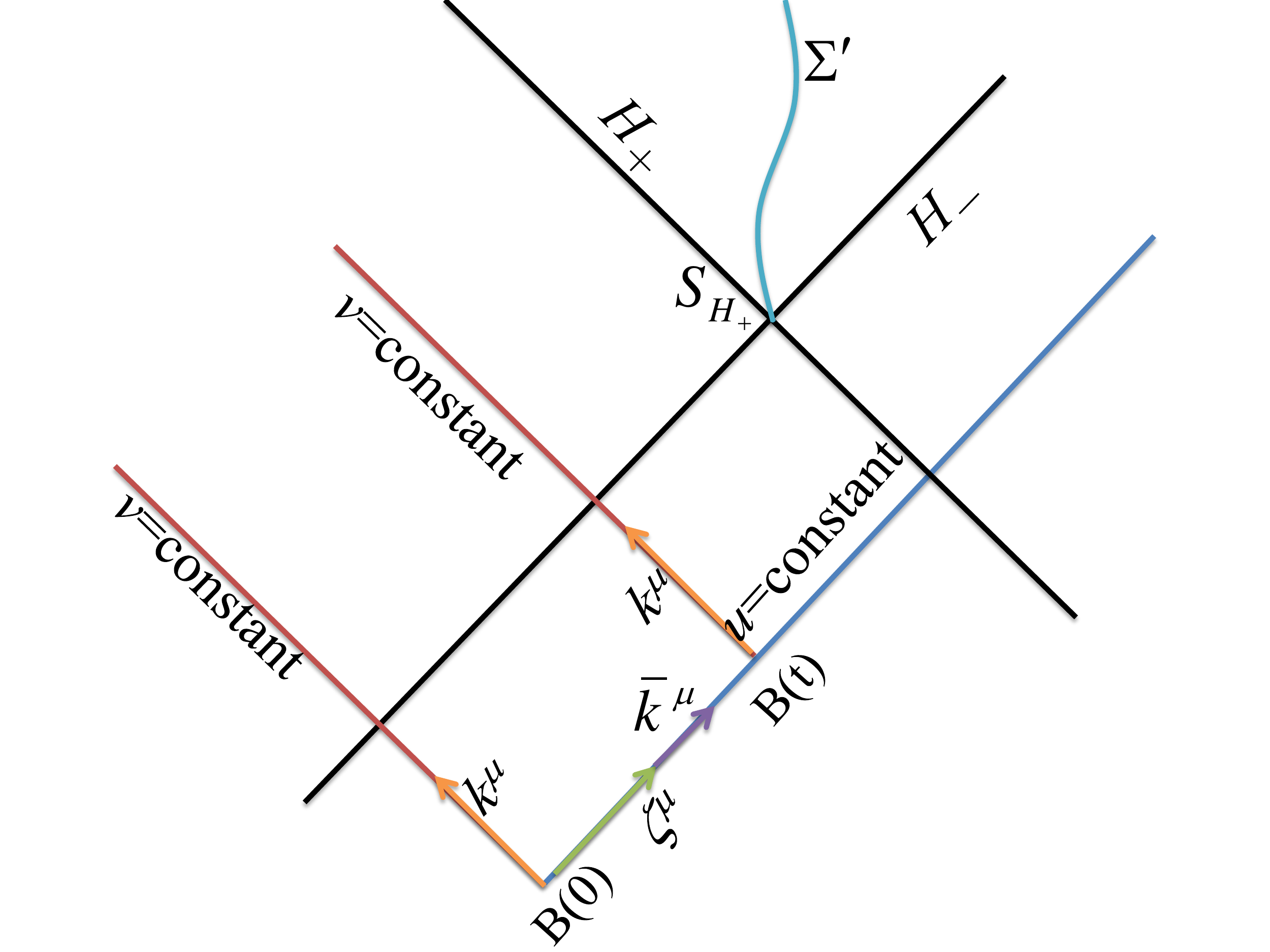}
  \caption{The joint terms in the WDW patch. $H_+$ and $H_-$ stand for future ONKH and past ONKH, respectively. $k^\mu$ is  affinely parameterized normal vector field of null surfaces characterized by $v=$costant. $\bar{k}^\mu$ is affinely parameterized normal vector field of null surfaces characterized by $u=$constant. $B(t)$ is the joint at the time $t_L=t$ and the  flow $B(0)\mapsto B(t)$ is generated by vector field $\zeta^\mu$. The surface  $u=$costant is fixed but the surface $v=$constant evolves with time $t$. At the late time limit, the suface $u=$costant coincidents with $H_-$. Hypersurface $\Sigma'$ stands for the timelike 3-surface in inner region of ONKH, which is orthogonal to $\xi^\mu$ and intersects with $H_\pm$ at $S_{H_+}$.}\label{Fig2}
\end{figure}

Now let's consider the contribute from joint term. For the case (b2), only the joint term which is formed by two past null sheets coming from time slices $t_R$ and $t_L$ has contribution on the action growth. Let two intersecting past null surfaces be characterized by two scalar fields $v,u$ such that $v=$constant and $u=$constant respectively (see Fig.~\ref{Fig2}).
Assume $k_\alpha=-c(\td v)_\alpha$ and $\bar{k}_\alpha=\bar{c}(\td u)_\alpha$ to be the two affinely parameterized dual normal vectors for them. Here $c$ and $\bar{c}$ are arbitrary positive constants. For given fixed right time $t_R$, the null sheet characterized by $u=$constant will be fixed. The null sheets given by $v=$constant will evolve with time $t_L$. For a specified initial time $t_L=0$, let's assume the joint locates at $B(0)$. Then for any time $t_L=t$, the joint will locates at $B(t)$. With increasing the time $t_L$, the joint $B(t)$ will flow along the null surface characterized by $u=$constant. The late time limit means that $t_R\rightarrow\infty$ and $u=$constant surface will meet the one branch of bifurcated Killing horizon. Based on the results in Ref. \cite{Lehner:2016vdi}, the contribution from the joint terms can be expressed as,
\begin{equation}\label{joints1a}
  \frac{\td }{\td t}\mathcal{A}_{\text{in,joint}}=\frac1{8\pi}\frac{\td }{\td t}\oint_{B(t)}a\td S=\frac1{8\pi}\oint_{B(t)}\mathcal{L}_{\zeta}(a\td S)\,.
\end{equation}
Here $\td S$ is the area element on $B(t)$, $a=\ln(-\bar{k}^\mu k_\mu/2)$ and $\zeta^\mu$ is the generator of flow $B(0)\mapsto B(t)$, $\mathcal{L}_\zeta$ is the Lie derivative operator with respective to vector field $\zeta^\mu$. First we see that,
\begin{equation}\label{Lied0a}
\begin{split}
  \bar{k}^\alpha\mathcal{L}_\zeta k_\alpha&=-c\bar{k}^\alpha\mathcal{L}_\zeta (\td v)_\alpha=-c\bar{k}^\alpha(\td \mathcal{L}_\zeta v)_\alpha\\
  &=-c\bar{k}^\alpha\partial_\alpha(\zeta^\mu\partial_\mu v)=-c\mathcal{L}_{\bar{k}}(\zeta^\mu\partial_\mu v)\,.
  \end{split}
\end{equation}
On then other hand, at the late time limit, the flow is tangent to $H_-$ and $\zeta^\mu\doteq\xi^\mu$ as $H_-$ is a Killing horizon. Here the notation $\doteq$ means that equality is correct only when the joint $B(t)$ approaches to the Killing horizon $H_{\pm}$ infinitely. At the Killing horizon $H_\pm$, we can find that $\exists \kappa_\pm$ such that $\xi^\mu\nabla_\mu\xi^\nu\doteq\kappa_\pm\xi^\nu$ and $\bar{k}^\mu\doteq e^{-\Gamma}\xi^\mu$ with $\mathcal{L}_\xi \Gamma\doteq\kappa_\pm$. Then one can prove that,
\begin{equation}\label{Lied1}
\mathcal{L}_{\bar{k}}(\zeta^\mu\partial_\mu v)\doteq e^{-\Gamma}\mathcal{L}_\xi(\zeta^\mu\partial_\mu v)=0,~~\text{and}~\mathcal{L}_\zeta\td S\doteq0.
\end{equation}
as $\xi^\mu$ is a Killing vector and
\begin{equation}\label{Lied2}
  k_\mu\mathcal{L}_\zeta \bar{k}^\mu\doteq k_\mu\mathcal{L}_\xi (e^{-\Gamma}\xi^\mu)\doteq-\kappa_- \bar{k}_\mu k^\mu.
\end{equation}
The first equality sign is because that $\zeta^\mu$ and $\bar{k}^\mu$ are both tangent to $H_-$. Now one can take \eqref{Lied1} and \eqref{Lied2} into \eqref{joints1a} and find,
\begin{equation}\label{joints1}
  \frac{\td }{\td t}\mathcal{A}_{\text{in,joint}}=-\frac1{8\pi}\frac{\td }{\td t}\oint_{B(t)}\kappa_-\td S
\end{equation}
in the late time limit. On the other hand, one can easy show that the Komar mass contained in the Killing horizon $H_+$ is,
\begin{equation}\label{KomarKH}
   m(H_+)=\frac1{4\pi}\frac{\td }{\td t}\oint_{S_{H_+}}\kappa_+\td S+\frac3{4\pi\el^2}\int_{\Sigma'}\xi^\mu \td V_\mu\,.
\end{equation}
Here $\td V_\mu=-\xi^\mu/\sqrt{\xi^\mu\xi_\mu} \td V$. Combine the Eqs. \eqref{joints1},\eqref{KomarKH} and note $\kappa_-=-\kappa_+$ as these is a bifurcate Killing horizon and $S_{H_+}$  is associated with $B(t)$ by time translation generated by $\xi^\mu$at the late time limit, we can obtain that,
\begin{equation}\label{joints2}
  \frac{\td }{\td t}\mathcal{A}_{\text{in,joint}}=\frac12m_H+\frac3{8\pi\el^2}\int_{\Sigma'}\sqrt{\xi^\mu\xi_\mu}\td V\,.
\end{equation}
%
%

After combining the expressions \eqref{bulkL},\eqref{bdS1} and \eqref{joints2}, we can obtain
\begin{equation}\label{asdtineq1}
  \frac{\td \mathcal{A}}{\td t}= \frac{\td }{\td t}\mathcal{A}_{\text{in,bulk}}+ \frac{\td }{\td t}\mathcal{A}_{\text{in,bd}}+ \frac{\td }{\td t}\mathcal{A}_{\text{in,joint}}= 2m_H.
\end{equation}
As condition \eqref{strongcd1} leads to $M\geq m_H$, one can immediately reach the expected inequality \eqref{growthbound2} and saturation appears when $(T_{\mu\nu}-\frac12Tg_{\mu\nu})\xi^\mu\xi^\nu=0$. The vacuum Schwarzschild black hole  satisfies this condition, so it is one case which can saturate the bound \eqref{growthbound2}.

\subsection{Proof for cases (b1) and (c)}\label{timenull}
\begin{figure}
  \centering
  \includegraphics[width=0.37\textwidth]{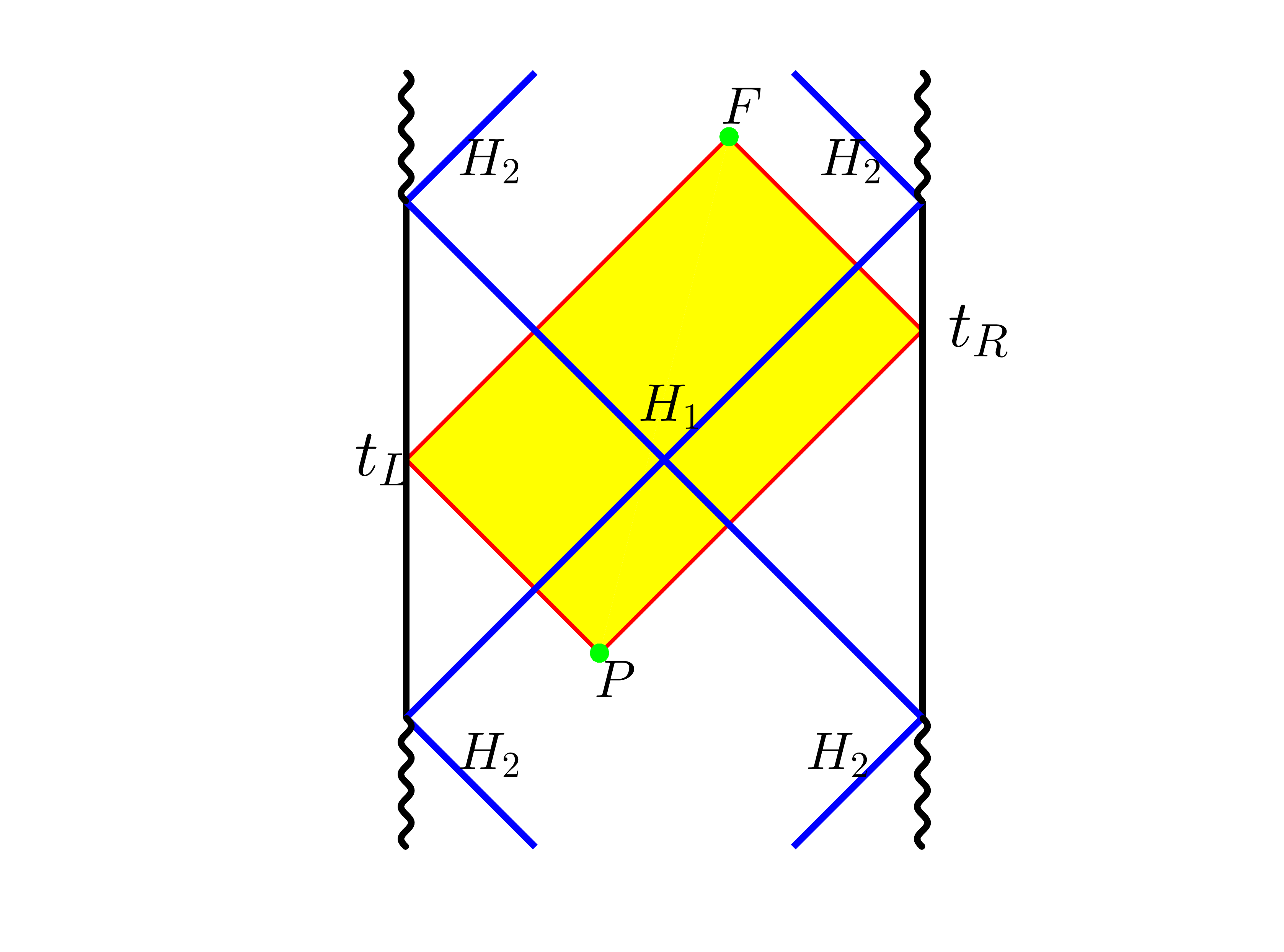}
  \caption{An example of WDW patch in a black hole with multiple nondegenerated Killing horizons. For given particular time-slices $t_L$ and $t_R$ at the two boundaries, the WDW patch is the yellow region with its boundary. $H_1$ is the ONKH and $H_2$ is the sub-ONKH. The codimension-2 surfaces $P$ and $F$ are the joints of two past and future null sheets coming from the boundary slices $t_L$ and $t_R$, respectively. At the late time limit $t_R\rightarrow\infty$, the null sheet from $t_R$ to P will coincide with one branch of the ONKH and the null sheet from the  $t_R$ to F will coincide with one branch of the sub-ONKH. }\label{timeWDW}
\end{figure}

\begin{figure}
  \centering
  \includegraphics[width=0.37\textwidth]{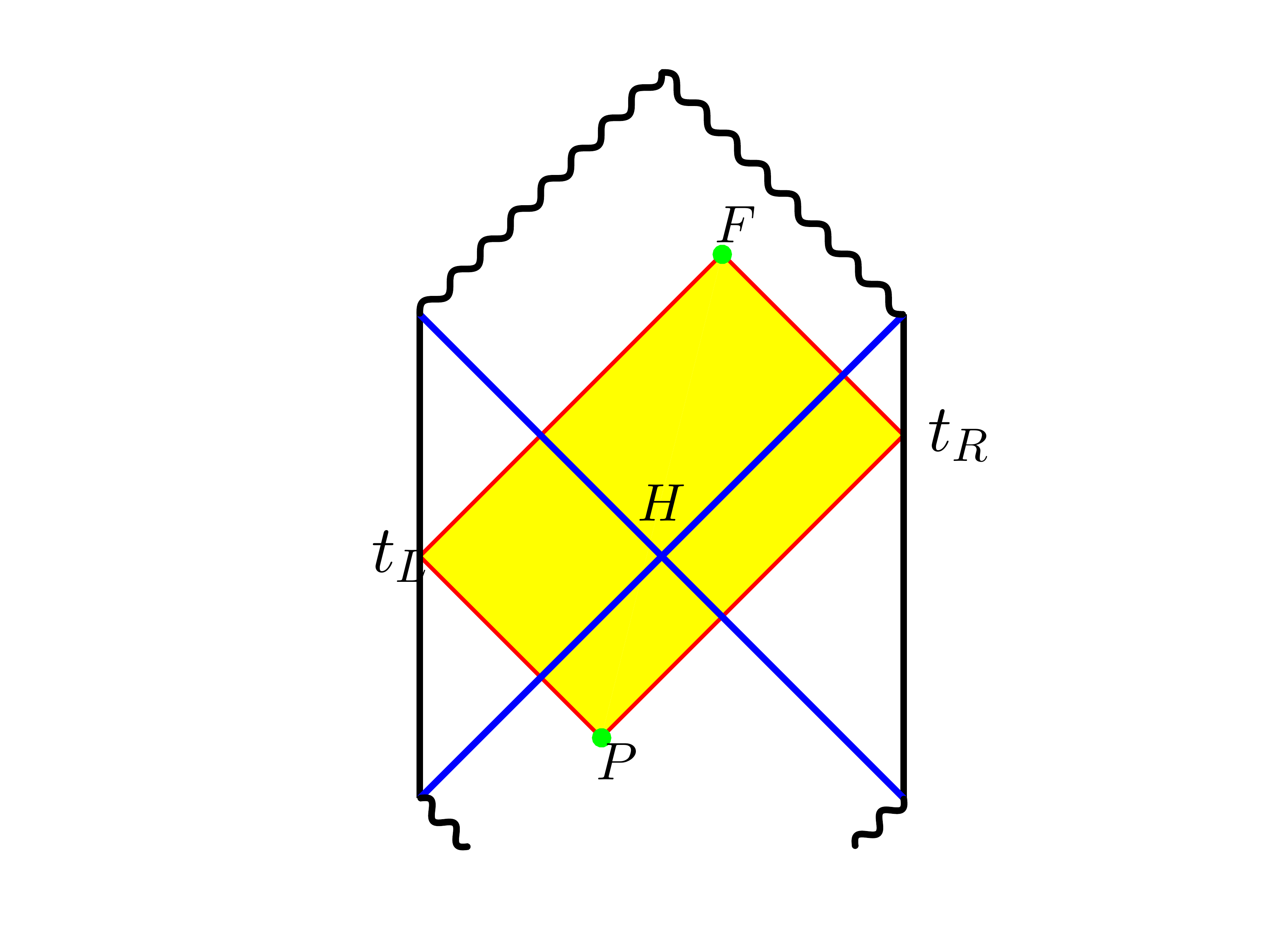}
  \caption{An example of WDW patch in a black hole with only one bifurcate Killing horizon and null singularity. For given particular time-slices $t_L$ and $t_R$ at the two boundaries, the WDW patch is the yellow region with its boundary. $H$ is the ONKH. The codimension-2 surfaces $P$ and $F$ are the joints of two past and future null sheets coming from the boundary slices $t_L$ and $t_R$, respectively. At the late time limit $t_R\rightarrow\infty$, the null sheet from $t_R$ to P will coincide with one branch of the ONKH and the null sheet from the  $t_R$ to F will coincide with the null singularity. }\label{nullWDW}
\end{figure}

Let's consider the cases (b1) and (c). For the case (b1), as $\xi^\mu$ must be tangent to the singularity, we can also build a local coordinate on the singularity and introduce the induced metric as the same as the form shown in the Eq.~\eqref{dsnull1}. Then the condition $\xi^2=0$ means $N=0$, so the metric on the singularity is degenerated and this singular surface must be null.  Similar to the case (b2), the action in the outside of the ONKH has no contribution on action growth in cases (b1) and (c). It is different from the case (b2) that the WDW patches in these two cases will not touch the singularities. This leads that there is no any boundary term need to be compute when we compute the action growth at the late time limit. Instead, a new additional null-null joint term will appear, which comes from intersection of two the future null sheets in the inner region of the ONKH, see the Figs.~\ref{timeWDW} and \ref{nullWDW} for examples.

Let $\mathcal{A}_{\text{in,bulk}}$, $\mathcal{A}_{\text{in,joint,F}}$ and $\mathcal{A}_{\text{in,joint,P}}$ stand for the contributions coming from the bulk integration, the joint term of past null sheets and the joint term of the future null sheets respectively. Then we can find,
\begin{equation}\label{dsdt2}
 \frac{\td \mathcal{A}}{\td t}= \frac{\td }{\td t}\mathcal{A}_{\text{in,bulk}}+ \frac{\td }{\td t}\mathcal{A}_{\text{in,joint,P}}+\frac{\td }{\td t}\mathcal{A}_{\text{in,joint,F}}\,.
 \end{equation}
Let's first consider the case (c). Similar to the case (b2), we can find that,
\begin{equation}\label{bulkL2}
  \frac{\td }{\td t}\mathcal{A}_{\text{in,bulk}}=-\frac3{8\pi\el^2}\int_{\Xi} \sqrt{\xi^\mu\xi_\mu}\td V\,.
\end{equation}
Here $\Xi$ is the timelike hypersurface which is orthogonal to $\xi^\mu$, and $\Xi$ connects the ONKH and the sub-ONKH.  Repeat the similar process from the Eq.~\eqref{joints1a} to Eq.~\eqref{joints2},  we can compute the contribution of joint forming by the future and past null sheets respectively. Then we can find,
\begin{equation}\label{joints4}
  \frac{\td }{\td t}\mathcal{A}_{\text{in,joint,P}}+\frac{\td }{\td t}\mathcal{A}_{\text{in,joint,F}}=m_H|^1_2+\frac3{8\pi\el^2}\int_{\Xi}\sqrt{\xi^\mu\xi_\mu}\td V\,.
\end{equation}
Here $m_H|^1_2:=m_H|_{H_1}-m_H|_{H_2}$. As the spacetime in between $H_1$ and $H_2$ is vacuum, we see $m_H|_{H_1}=m_H|_{H_2}$. By the strong energy condition and the condition $M\geq0$, we see,
\begin{equation}\label{dsdtc1}
 \frac{\td \mathcal{A}}{\td t}= 0\leq2M.
\end{equation}
Hence, the bound equation is satisfied.

For the case (b1), though the null singularity replaces the sub-ONKH, the computation about the joint term at the joint $P$ can still be used into the joint term in $F$. This is because that the null singularity is also a Killing horizon if we add such null surface into the space-time and we don't need to really touch the null singular surface when we compute the joint $F$. This shows that we still can obtain the Eq.~\eqref{dsdtc1} and the growth bound is still correct for the case (b1).

\section{conclusion and discussion}
To conclude, this paper studies a new universal inequality about the action growth rate in eternal black holes. For the cases that matter fields are restricted in the outside of horizons and decay rapidly near the AdS boundary, it shows that strong energy condition is a sufficient condition to insure the bound inequality~\eqref{growthbound2} and vacuum  black hole is one (not the only one) of which have fastest computational complexity growth if the space-time satisfies some general conditions. Although the inequality is proposed by the consideration of CA conjecture, the proof of it in this paper doesn't rely on the correctness of CA conjecture and holographic duality. The proofs here don't involve the details of matter fields, however, some strong conditions are still involved. Let's make some discussions on possibility of weakening these requirements.

Two strong requirements are that the space-time should be static and matter fields appear in the outside of ONKH. The main reason for these conditions is that by them we can convert the action growth rate bound inequality into a geometrical formulation. For the cases that space-time does not contain timelike (at the boundary) Killing vector field or matter fields can extend into the inner region of ONKH, the action of matter fields has contributions on the total action growth rate. As the result, we also have to compute the bulk integration for matter Lagrangian which in general cases cannot be written as any geometrical formulation and energy condition can not give enough restriction to insure the action growth rate bound. For example, in the static space-time which satisfies the case (b2) but matter field only appear in the inner region of ONKH, then in the Eq.~\eqref{dsdt1}, the bulk term should be replaced by
\begin{equation}\label{matterA1}
  \frac{\td}{\td t}\mathcal{A}_{\text{bd}}=\frac1{16\pi}\int_{\Sigma'} \left(\mathcal{L}_m-\frac{T}2-\frac6{\el^2}\right)\sqrt{\xi^\mu\xi_\mu}\td V\,.
\end{equation}
Here $\mathcal{L}_m$ is the Lagrangian of matter fields. Strong energy condition cannot give the enough restriction on the integration about $\mathcal{L}_m-T/2$, so the action growth rate bound can be break if we choose $\mathcal{L}_m$ suitably.

However, it still has possibility to weaken the static space-time into only stationary space-time. The most important example is the vacuum rotational black hole. For the vacuum axisymmetric rotational black, the metric can be given by Kerr-AdS solution, and the action growth rate has been given by the Ref.~\cite{Cai:2016xho} for 4 dimensional case. It turns out that the action growth rate can be expressed as,
\begin{equation}\label{KerrAdS}
  \frac{\td\mathcal{A}}{\td t}=[M-\Omega(r) J]|^{r_+}_{r_-}\,.
\end{equation}
Here $M$ is the total energy, $J$ is the total angular momentum and $\Omega(r)$ is the angular velocity. $r_{\pm}$ means the outer and inner event horizons. The expression~\eqref{KerrAdS} is also correct for BTZ black hole. After some algebras, one can directly check that $\td\mathcal{A}/\td t<2M$, so the bound equation Eq.~\eqref{growthbound2} is still correct. This means if the rotational black hole is a Kerr-AdS$_4$ black hole, the action growth rate bound is also true. In fact, a similar result can be obtained for all single parameter Kerr-AdS$_d$ black holes when dimension $d\geq3$~\cite{Huang:2016fks}.  By some algebras, one can show that we still have $\td\mathcal{A}/\td t<2M$. However, it is not clear if the bound inequality is still correct for multiple axes Kerr-AdS black holes in the higher dimension cases. This is also an interesting problem and worth to studying in the future. If the rotational black hole doesn't have axisymmetry or some general matter field appears, it's not very clear about the structures of the horizons and the property of the singularities. It is important and interesting to study if the inequality~\eqref{growthbound2} can obtain a general proof under some universal conditions in this case.

The other reason that the matter fields are restricted outside is that if some general matter fields can extend into the inner region of the black hole, then the matters may touch the singularity and may lead some divergences in the action of matter fields. At present, it's not clear that if the on-shell action has a finite value when some general matter fields touch the singularity. However, for some special matter fields which have clear Lagrangian and the systems have good symmetries such as spherical or planar symmetry, the action growth rates can be compute directly by analytical or numerical solutions. For example, RN-AdS$_d$ black hole has been studied by Refs.~\cite{Brown:2015bva,Brown:2015lvg,Lehner:2016vdi,Cai:2016xho} and is found that,
\begin{equation}\label{RND}
  \frac{\td\mathcal{A}}{\td t}=[M-\mu(r) Q]|^{r_+}_{r_-}\,.
\end{equation}
Here $M$ is the total energy, $Q$ is the total charge and $\mu(r)$ is the chemical potential. $r_{\pm}$ means the outer and inner event horizons. After some algebras,  one can also check that Eq.~\eqref{RND} satisfies the bound Eq.~\eqref{growthbound2} in any dimension which is larger than 2.  It is also very interesting to check other special matter fields such scalar field, SU(N) field, $p$-form field and so on. One recent work about this problem is Ref.~\cite{Cai:2017sjv}, which considers two non-trivial matter fields which extend into the horizon. One is the charged dilaton field and the other is the Born-Infeld field. Ref.~\cite{Cai:2017sjv} finds the action growth rate bound is still satisfied. An other interesting thing in Ref.~\cite{Cai:2017sjv} is that it considers a phantom Maxwell field, which breaks the strong energy condition and also breaks the action growth rate bound as expected. Then the results in Ref.~\cite{Cai:2017sjv} seems to show that even in the case that matter field can extend into the horizon, the action growth rate bound Eq.~\eqref{growthbound2} is still correct for large classes of physical interesting matter fields and strong energy condition still plays important role.  Overall, in very general cases that the systems are lack of symmetries, it's not clear if the action growth rate at the late time limit is well defined when matter fields extend into the horizon. This makes this paper to restrict the matter fields outside.

The condition on the Killing horizon, i.e., the condition on $\mathcal{K} $, though it is not satisfied for all the space-time, it can cover the most of interesting cases in physics. This condition simplifies the topology of WDW patch. Under this restriction, the space-time has three cases: $\mathcal{K} =\emptyset$, $\mathcal{K} =\{H_o\}$ and $\{H_o,H_{\text{sub-o}}\}\subset \mathcal{K}$. The cases  $\mathcal{K} =\emptyset$ and $\{H_o,H_{\text{sub-o}}\}\subset \mathcal{K}$ has been covered by the cases (a) and (c) in the section~\ref{gencase}. The special interesting should be paid to the conditions on the cases (b2). Because of the restriction of topology, the conditions (b1)and (b2) cannot covered all the cases of $\mathcal{K} =\{H_o\}$. For example, the planar and hyperbolic AdS black  are belong to the case $\mathcal{K} =\{H_o\}$  but not covered by conditions (b1) and (b2). However, by the discussions about spherical Schwarzschild AdS black in the section~\ref{spher}, one can easy see that the action growth rate bound is still correct for these two cases if they are static and matter field only appear in the outside of the horizon. It is very interesting to study if we can find the proof which can cover all the cases of  $\mathcal{K} =\{H_o\}$ in the future study. In the cases(a), (b1) and (c), the additional condition $M\geq0$ cannot be obtained only by strong energy condition. One can use dominant energy condition to replace it. In the cases (a), (b1) and (c), the requirement $M\geq0$ is necessary. If the total mass $M<0$, the action growth rate bound will be broken.

This paper focused on the four dimensional black holes, but the method can also be generalized into other dimensions. The assumption on $\mathcal{K} $ can obtain well meaning for any dimensional black hole. Then the discussion for the cases (a), (b1) and (c) can directly be applied into these dimensions. The proofs in the subsections~\ref{b2bulk} and \ref{b2joint} have no any difficult when they are applied into general higher dimensions. Then we can find that the Eqs.~\eqref{bulkL} and \eqref{joints2} become,
\begin{equation}\label{bulkLd}
  \frac{\td }{\td t}\mathcal{A}_{\text{in,bulk}}=-\frac{(d-1)(d-2)}{16\pi\el^2}\int_{\Sigma'} \sqrt{\xi^\mu\xi_\mu}\td V\,,
\end{equation}
and,
\begin{equation}\label{joints2d}
  \frac{\td }{\td t}\mathcal{A}_{\text{in,joint}}=\frac{d-3}{d-2}m_H+\frac{(d-1)(d-2)}{16\pi\el^2}\int_{\Sigma'}\sqrt{\xi^\mu\xi_\mu}\td V\,.
\end{equation}
These show that for higher dimension black hole, if it satisfies the conditions in the cases (b1) and (c), then the action growth rate bound is still satisfied. The case (b2) is the only one that should be considered specially. For the case that dimension $d>4$, the topology constraint  can be generalized as $R\times S^{d-2}$. This makes the proof in the subsection~\ref{b2s} is subtle. For the general dimension $d>4$, following the process from \eqref{decomp1} to \eqref{bdS0c}, we can obtain that,
\begin{equation}\label{bdS0cd}
\begin{split}
  &\frac{\td }{\td t}\mathcal{A}_{\text{in,bd}}(r)-\frac{d-1}{d-2}m_H\\
  =&\frac1{8\pi}\oint(\frac{d-2}{d-3}\sqrt{h}\partial_rN^2+ N^2\partial_r\sqrt{h})N^{-1}\td^{d-2}x\\
  &+\frac{(d-2)(d-1)^2}{16\pi\el^2}\int_{\Sigma_{r<\epsilon}} N\sqrt{h}\td r\td^{d-2}x\,.
  \end{split}
\end{equation}
For even dimension, we can use generalized Gauss-Bonnet theorem to replace the topology constrain $\oint\mathcal{R}^{(2)}\sqrt{h}\td^2x=8\pi$ and show that the leading term of the metric in the vicinity of the singularity must have following form,
\begin{equation}\label{decompd}
  \td s^2=-\td r^2+r^{2p_t}f_1(\vec{x})\td t^2+r^{2p_0}\tilde{h}_{AB}(\vec{x})\td x^A\td x^B\,.
\end{equation}
Here $x^A$ with $A=1,2,\dots,d-2$ is the coordinate laying on the singular surface. For the general dimension $d>4$, $\tilde{h}_{AB}(\vec{x})$ is still smooth function of $x^A$.  Then taking them into Einstein's equation and following the similar analysis in the appendix~\ref{app3}, one can finally find that $p_t=-(d-3)/(d-1)$ and $p_0=2/(d-1)$. This leads,
\begin{equation}\label{bdS1d}
  \frac{\td }{\td t}\mathcal{A}_{\text{in,bd}}=\frac{d-1}{d-2}m_H\,,
\end{equation}
so we still find that action growth rate is $2m_H$. This shows that the case (b2) can insure the bound Eq.~\eqref{growthbound2} in higher even dimension. We also see that $p_t$ and $p_0$ still satisfy the Kasner's relationship: $p_t+(d-2)p_0=p_t^2+(d-2)p_0^2=1$. However, for odd dimension, the topology $R\times S^{d-2}$ cannot give metric such as Eq.~\eqref{decompd}, so the case (b2) cannot directly generalized into the odd dimension. This is also a research direction in the future.

In this paper, we see that strong energy condition is a sufficient condition, which is because what we need is that the Komar mass at the AdS boundary is not less than its value at the ONKH. Strong energy condition is a sufficient condition. Can we weaken such condition? Especially, when we consider the facts that strong energy condition contain the null energy condition and many matter fields can break strong energy condition, it is very interesting to investigate how to weaken this condition. However, if we restrict the candidates only including four local energy conditions which are strong, weak, null and dominant energy conditions, then only strong energy condition is the one which can insure the action growth rate bound without adding more additional assumptions than what we have done in the cases (a), (b1), (b2) and (c). For example, for the case (b2), the action growth rate is just $2m_H$. Then without additional assumptions on the symmetry or matter fields, only strong energy condition can insure the total mass $M$ is not less than $m_H$.

\acknowledgments{I would thank  Adam R. Brown, Liming Cao and Shaojiang Wang for helpful discussions.}

\appendix
\section{Bulk integration term}\label{app1}
In this appendix, I will show that the bulk integration term in the action growth rate can be written into Eq. \eqref{bulkL}. At the inner region of ONKH, as the matter fields are zero, the bulk action of the system is,
\begin{equation}\label{bulkact}
\begin{split}
  \mathcal{A}_{\text{in,bulk}}&=\frac1{16\pi}\int_\mathcal{V}(R+6/\el^2)\sqrt{-g}\td^4x\\
  &=-\frac3{8\pi\el^2}\int_\mathcal{V}\sqrt{-g}\td^4x\,.
  \end{split}
\end{equation}
Here $\mathcal{V}$ is the inner region of Killing horizon, which is given by $\xi^2=\xi^\mu\xi_\mu>0$.

Firstly, we introduce a time orthogonal coordinate $\{x^\mu\}=\{t,x^1,x^2,x^3\}$ so that the line element in the inner region of ONKH is,
\begin{equation}\label{metricinner}
  \td s^2=g_{\mu\nu}\td x^\mu\td x^\nu=N^2\td t^2+\gamma_{ij}\td x^i\td x^j\,.
\end{equation}
Here $N^2=\xi^\mu\xi_\mu$, $\xi^\mu=(\partial/\partial t)^\mu$ and $i,j=1,2,3$. In this coordinate the right hand of Eq. \eqref{bulkL} can be expressed as,
\begin{equation}\label{righthandbulk}
  \frac3{8\pi\el^2}\int_{\Sigma'} \sqrt{\xi^\mu\xi_\mu}\td V=\frac3{8\pi\el^2}\int_{\xi^2>0} N\sqrt{-\gamma}\td^3x\,.
\end{equation}
But the coordinate $\{x^\mu\}$ isn't suitable to compute the growth rate of $\mathcal{A}_{\text{in,bulk}}$ as the integration region is bounded by null surfaces. In order to compute its value, let's introduce a null time coordinate $\tilde{u}$, which is given by,
\begin{equation}\label{nullu1}
  (\td \tilde{u})_\mu=(\td t)_\mu+J_i(\td x^i)_\mu
\end{equation}
with three spatial functions $J_i=J_i(\vec{x})$ and
\begin{equation}\label{nullu1}
  g^{\mu\nu}(\td \tilde{u})_\mu(\td \tilde{u})_\nu=0\,.
\end{equation}
Such null coordinate always exists. In general, a null time coordinate $\tilde{u}$ can always be defined by,
\begin{equation}\label{nullu1}
  \td \tilde{u}=J_0\td t+J_i\td x^i
\end{equation}
with some functions $J_0,J_1,J_2$ and $J_3$. As the $(\partial/\partial t)^\mu$ is Killing vector field, these four functions are all independent of $t$. Then integrable condition $\td(\td\tilde{u})=0$ shows that,
\begin{equation}\label{conditionsu1}
  \partial_iJ_0=0,~~\partial_iJ_l-\partial_lJ_i=0\,.
\end{equation}
This means that $J_0$ is a constant so we can set $J_0=1$ and $J_i=\partial_i\Phi(\vec{x})$ for a scalar field $\Phi(\vec{x})$. The null condition shows,
\begin{equation}\label{conditionsu2}
  1/N^2+\gamma^{ij}\partial_i\Phi\partial_j\Phi=0\,.
\end{equation}
Here $\gamma^{ij}$ is the inverse of $\gamma_{ij}$. We see that the solution of Eq. \eqref{conditionsu2} always exist and $\tilde{u}=t+\Phi(\vec{x})$. We can choose the sign of $\Phi(\vec{x})$ so that $\tilde{u}=$constnat gives the infalling null surface.\footnote{If the $(\td u)_\mu$ is the affine dual normal vector, i.e., $g^{\alpha\beta}(\td u)_\beta\nabla_\alpha(\td u)_\mu=0$, then such scalar function $\tilde{u}$ is just the scalar field $u$ shown in the Fig. \ref{Fig2}.}

Now in the null coordinate $\{\tilde{x}^\mu\}=\{\tilde{u},\vec{x}\}$, the null boundary of WDW patch in the inner region of ONKH is just a $\tilde{u}=$cosntant surface. The line element then is,
\begin{equation}\label{metricux}
\begin{split}
  \td s^2&=\tilde{g}_{\mu\nu}\td\tilde{x}^\mu\td\tilde{x}^\nu\\
  &=N^2\td\tilde{u}^2-2N^2J_i\td x^i\td\tilde{u}+\tilde{\gamma}_{ij}\td x^i\td x^j.
  \end{split}
\end{equation}
with $\tilde{\gamma}_{ij}=\gamma_{ij}+N^2J_iJ_j$. One can directly check that the determinant of metrics in the coordinates $\{t,\vec{x}\}$ and $\{\tilde{u},\vec{x}\}$ are the same , i.e., $\sqrt{-\tilde{g}}=\sqrt{-g}=N\sqrt{-\gamma}$. This result can also be obtained by following way. One can first compute the inverse metric in the  coordinate $\{\tilde{x}^\mu\}=\{\tilde{u},\vec{x}\}$ by coordinate transformation formula $\tilde{g}^{\mu\nu}=(\partial\tilde{x}^\mu/\partial x^\alpha)(\partial\tilde{x}^\nu/\partial x^\beta)g^{\alpha\beta}$, which gives,
\begin{equation}\label{invesg2}
  \tilde{g}^{00}=1/N^2+\gamma^{ij}J_iJ_j,~~\tilde{g}^{0i}=\gamma^{ij}J_j,~~\tilde{g}^{ij}=\gamma^{ij}\,.
\end{equation}
Then using the formula $\td\tilde{g}/\td J_i=\tilde{g}\tilde{g}^{\mu\nu}(\td\tilde{g}_{\mu\nu}/\td J_i)$, one can find that,
\begin{equation}\label{dgdJi}
  \tilde{g}^{-1}\frac{\td\tilde{g}}{\td J_i}=-\tilde{g}^{0i}N^2+\tilde{g}^{ij}J_jN^2=0\,.
\end{equation}
This equation means that $\tilde{g}$ is independent of $J_i$, so we have $g=\tilde{g}$.

In the coordinate $\{\tilde{u},\vec{x}\}$, the variation of $\mathcal{A}_{\text{in,bulk}}$ from the time $t$ to $t+\delta t$ equals to its variation from $\tilde{u}$ to $\tilde{u}+\delta t$, i.e.,
\begin{equation}\label{bulkint0}
  \delta\mathcal{A}_{\text{in,bulk}}=-\frac3{8\pi\el^2}\int_{\tilde{u}}^{\tilde{u}+\delta t}\td \tilde{u}\int_{\xi^2>0} \sqrt{-\tilde{g}}\td^3x\,.
\end{equation}
Then we obtain the action growth rate coming from the inner bulk integration,
\begin{equation}\label{bulkint1}
\begin{split}
  \frac{\td }{\td t}\mathcal{A}_{\text{in,bulk}}&=-\frac3{8\pi\el^2}\int_{\xi^2>0} \sqrt{-\tilde{g}}\td^3x\\
  &=-\frac3{8\pi\el^2}\int_{\xi^2>0} N\sqrt{-\gamma}\td^3x\,.
  \end{split}
\end{equation}

By comparing the Eqs. \eqref{bulkint1} and \eqref{righthandbulk}, one can immediately obtain the Eq. \eqref{bulkL}.

\section{Boundary contribution near the singularity}\label{app2}
For the surface $r=\epsilon$ near the singularity, the unit future/outward directed normal vector is $r^\mu=[0,-1,0,0]$, so the trace of extrinsic curvature is,
\begin{equation}\label{extriK}
  K=\nabla_\mu r^\mu=-\frac1{N\sqrt{|h|}}\partial_r(N\sqrt{|h|}).
\end{equation}
and the volume element $\td V=N\sqrt{|h|}\td t\td^2x$. Then we can get the action growth rate coming from the boundary term, which is,
\begin{equation}\label{bdS0as}
  \frac{\td }{\td t}\mathcal{A}_{\text{in,bd}}(r)=\frac{\varepsilon}{8\pi}\oint \partial_r(N\sqrt{|h|})\td^2x.
\end{equation}

Before we give the details about Eq.~\eqref{bdS0b}, let's first clarify the convention of directed volume element and surface element. In order to match the Komar mass defined in Eq.~\eqref{Smarr2}, the direction for directed volume $\td V_\nu=n_\mu\td V$ is appointed as: $n_\mu$ is future directed when $n_\mu$ is timelike and is inward directed when $n_\mu$ is spacelike. By this convention for directed volume element, the Gauss theorem for a surface integration is,
\begin{equation}\label{GaussS}
 \frac12 \oint_{\partial\Sigma}F^{\mu\nu}\td S_{\mu\nu}=\int_{\Sigma}\nabla_\mu F^{\mu\nu}\td V_\nu
\end{equation}
for any antisymmetric tensor field $F^{\mu\nu}$. The directed surface element is defined as $\td S_{\mu\nu}=-2\varepsilon n_{[\mu}l_{\nu]}\td S$. Here $\td S=\sqrt{|h|}\td^2x$ and $l^\nu$ is unit vector which is tangent to $\Sigma$ and normal to $\partial\Sigma$, and it is inward direction if it's spacelike and future directed if it's timelike.\footnote{Another common direction appointment is that $n_\mu$ is past directed when $n_\mu$ is timelike and $n_\mu$ is outward directed when $n_\mu$ is spacelike. This lead to a ``$-$" sign in the volume integration of Eq. \eqref{GaussS}.}

The Komar mass contained by the surface of $r=\epsilon$ is expressed by following integration,
\begin{equation}\label{Komars1}
\begin{split}
  m(r)&=-\frac1{8\pi}\oint[\frac12(\td\xi)_{\mu\nu}-\frac{3\omega_{\mu\nu}}{\el^2}]\td S^{\mu\nu}\\
  &=-\frac1{16\pi}\oint(\td\xi)_{\mu\nu}\td S^{\mu\nu}+\frac3{4\pi\el^2}\int_{\Sigma_{(r<\epsilon)}}\xi^\mu\td V_\mu
  \end{split}
\end{equation}
$\td V_\mu=n_\mu\td V$ is the vector volume on $\Sigma_{(r<\epsilon)}$ with $n_\mu=-\xi_\mu N^{-1}$. $\td S_{\mu\nu}$ is the directed surface element at the two-surface by fixing $t$ and $r=\epsilon$. One can find that $\td S_{\mu\nu}=-2\varepsilon n_{[\mu}r_{\nu]}\td S$. Note that,
\begin{equation}\label{Komars2}
\begin{split}
&\frac12(\td\xi)_{\mu\nu}\td S^{\mu\nu}=-\varepsilon(\td\xi)_{\mu\nu}n^\mu r^\nu\td S\\
&=\varepsilon N^{-1}r^\nu\xi^\mu(\partial_\mu\xi_\nu-\partial_\nu\xi_\mu)\td S\\
&=-\varepsilon N^{-1}r^\nu\xi^\mu\partial_\nu\xi_\mu\td S\\
&=-\varepsilon N^{-1}r^\nu\partial_\nu(\xi^\mu\xi_\mu)\td S\\
&(\because \xi^\nu=[1,0,0,0]\Rightarrow\partial_\mu\xi^\nu=0)\\
&=\varepsilon N^{-1}\partial_rN^2\td S
\end{split}
\end{equation}
we can find that Komar mass can be expressed as Eq.~\eqref{bdS0b}

\section{Asymptotic behavior near the singularity}\label{app3}
This appendix will show how to analyze the behavior of metric near the singularity. With the forms  $N=r^{p_t}f_1(\theta,\phi)$ and $h_{AB}=r^{2p_{AB}}\tilde{h}_{AB}(\theta,\phi)$(no summation between $2p_{AB}$ and $\tilde{h}_{AB}$), the line element \eqref{decomp3} becomes,
\begin{equation}\label{decomp3a}
  \td s^2=-\td r^2+r^{2p_t}f_1^2\td t^2+r^{2p_{AB}}\tilde{h}_{AB}\td x^A\td x^B.
\end{equation}
Let us first show that $p_{AB}=p_0$. For any 2-surface by fixing $t$ and $r$, the line element in this closed two surface is,
\begin{equation}\label{decomp3a}
  \td s^2_{(2)}=r^{2p_{\theta\theta}}\tilde{h}_{\theta\theta}\td\theta^2+2r^{2p_{\theta\phi}}\tilde{h}_{\theta\phi}\td\theta\td\phi+r^{2p_{\phi\phi}}\tilde{h}_{\phi\phi}\td\phi^2
\end{equation}
If $p_{\theta\theta}, p_{\theta\phi}$ and $p_{\phi\phi}$ are all zeros, line element \eqref{decomp3a} describes a regular closed surface which is topological homeomorphism to a unit sphere. For the case that any one of them is nonzero, let's find the restriction from the topology. As when we compute the scalar curvature $\mathcal{R}^{(2)}$, only the derivatives with respect to $\{\theta,\phi\}$ are involved, after some algebras, one can find that the curvature has following form,
\begin{equation}\label{scalarR}
\begin{split}
  \mathcal{R}^{(2)}&=\frac1{h^2}[r^{2(p_{\theta\theta}+ p_{\theta\phi}+p_{\phi\phi})}F_1(\theta,\phi)+r^{4p_{\theta\theta}+ 2p_{\theta\phi}}F_2(\theta,\phi)\\
 &+r^{4p_{\theta\theta}+2p_{\phi\phi}}F_3(\theta,\phi)+r^{4p_{\theta\phi}+2p_{\theta\theta}}F_4(\theta,\phi)\\
 &+r^{4p_{\theta\phi}+2p_{\phi\phi}}F_5(\theta,\phi)+r^{4p_{\phi\phi}+2p_{\theta\theta}}F_6(\theta,\phi)\\
 &+r^{4p_{\phi\phi}+2p_{\theta\phi}}F_7(\theta,\phi)+r^{6p_{\theta\phi}}F_8(\theta,\phi)
 \end{split}
\end{equation}
and,
\begin{equation}\label{deth1}
  h=r^{2(p_{\theta\theta}+p_{\phi\phi})}\tilde{h}_{\theta\theta}\tilde{h}_{\phi\phi}-r^{4p_{\theta\phi}}\tilde{h}_{\theta\phi}^2.
\end{equation}
Here $F_i$ with $i=1\cdots8$ are smooth functions. As the 2-surface is spacelike, we have $h>0$. This means that $p_{\theta\theta}+p_{\phi\phi}\leq2p_{\theta\phi}$, and when $r\rightarrow0$, we have
\begin{equation}\label{deth1}
  h\rightarrow r^{2(p_{\theta\theta}+p_{\phi\phi})}h^*
\end{equation}
Here $h^*=\tilde{h}_{\theta\theta}\tilde{h}_{\phi\phi}$ if $p_{\theta\theta}+p_{\phi\phi}<2p_{\theta\phi}$ and $h^*=\tilde{h}_{\theta\theta}\tilde{h}_{\phi\phi}-\tilde{h}_{\theta\phi}^2$ if $p_{\theta\theta}+p_{\phi\phi}=2p_{\theta\phi}$. This shows that,
\begin{equation}\label{intR2}
\begin{split}
  8\pi=&\oint\td^2x\frac1{h^{*3/2}}[r^{2p_{\theta\phi}-(p_{\theta\theta}+p_{\phi\phi})}F_1(\theta,\phi)\\
  &+r^{p_{\theta\theta}+ 2p_{\theta\phi}-3p_{\phi\phi}}F_2(\theta,\phi)+r^{p_{\theta\theta}-p_{\phi\phi}}F_3(\theta,\phi)\\
 &+r^{4p_{\theta\phi}-p_{\theta\theta}-3p_{\phi\phi}}F_4(\theta,\phi)+r^{4p_{\theta\phi}-p_{\phi\phi}-3p_{\theta\theta}}F_5(\theta,\phi)\\
 &+r^{p_{\phi\phi}-p_{\theta\theta}}F_6(\theta,\phi)+r^{p_{\phi\phi}+2p_{\theta\phi}-3p_{\theta\theta}}F_7(\theta,\phi)\\
 &+r^{6p_{\theta\phi}-3(p_{\theta\theta}+p_{\phi\phi})}F_8(\theta,\phi)]
  \end{split}
\end{equation}
It is easy to see that this equation is correct only when $p_{\theta\theta}=p_{\phi\phi}=p_{\theta\phi}=p_0$. Then line element \eqref{decomp3a} becomes,
\begin{equation}\label{decomp32}
  \td s^2_{(2)}=r^{2p_0}\td s^2_{S^2}.
\end{equation}
Here $\td s^2_{S^2}$ describe a smooth closed 2-surface, which is topological homeomorphism to a unit sphere. The Eq. \eqref{decomp32} has included the case $p_{AB}=0$. As any smooth closed surface which is topological homeomorphism to the unit sphere is conformal to the unit sphere, there is a coordinate on this 2-surface such that $\td s^2_{S^2}=f_2(\theta,\phi)\td\Omega^2$ with a smooth function $f_2(\theta,\phi)$ and standard line element $\td\Omega^2$ in the unit sphere. Finally, we obtain the line element shown in Eq.~\eqref{decomp4}.

To find the values of $p_t$ and $p_0$, we need to use Einstein's equation. With a cosmological constant, the vacuum Einstein's equation shows,
\begin{equation}\label{Eins1}
  {R^\mu}_{\nu}=-\frac{3}{\el^2}\delta^\mu_\nu
\end{equation}
Now we take the line element \eqref{decomp4} into Eq. \eqref{Eins1}. As the line element in Eq.~\eqref{decomp4} only contains the leading terms of real metric, two sides of \eqref{Eins1} should match each other at the leading order. This means that the coefficients of leading divergent terms in  ${R^\mu}_{\nu}$ are zeros.  Note that $p_0$ and $p_t$ can be the function of $r$. One can check that,
\begin{equation}\label{Einstr}
  {R^r}_A=\frac{\partial_A f_1}{2rf_1}[(p_0'-p_t')r\ln r+(p_0-p_t)]
\end{equation}
Here ``$'$" means the partial derivative with respect to $r$ and $A=\theta,\phi$. Then there are two different cases: $p_0(r)=p_t(r)$ or $f_1(\theta,\phi)$ is a constant.

If $p_0(r)=p_t(r)$ but $f_1(\theta,\phi)$ isn't a constant, then the leading term line element is,
\begin{equation}\label{decomp5a}
  \td s^2=-\td r^2+r^{2p_t}[f_1(\theta,\phi)\td t^2+f_2(\theta,\phi)\td\Omega^2].
\end{equation}
One can check that the leading term of ${R^r}_{r}$ is
\begin{equation}\label{Riccirr1}
  3{p_0'}^2(\ln r)^2+3\left(\frac{2p_0'p_0}{r}+p_0''\right)\ln r+\frac{6p_0'}r+\frac{3p_0(p_0-1)}{r^2}
\end{equation}
The leading divergent term should be canceled by suitable choice of $p_0(r)$. When $r\rightarrow0$, if $|p_0'|$ diverges more rapidly than $1/r$, one can see that the leading divergent term cannot be canceled no matter how to choose function $p_0(r)$. If $|p_0'|$ diverges more slowly than $1/r$ or as the same order as $1/r$, then the leading divergent terms in the expression \eqref{Riccirr1} can disappear only when,
\begin{equation}\label{p0values}
  p_0=0~~\text{or}~p_0=1.
\end{equation}
One easily see that these values of $p_0$ lead to line element \eqref{decomp5a} not having a singularity at $r=0$.

%
If $f_1(\theta,\phi)$ is a constant, then one can use the fact that ${R^t}_t, {R^r}_r$ and ${R^\theta}_\theta$ are all finite when $r\rightarrow0$ to show that, the leading divergent terms must be $1/r^2$ terms and the coefficients of these terms should be zeros by suitable choices of $p_t(r)$ and $p_0(r)$, which means following:
\begin{equation}\label{eqptp0}
  p_0(2p_0+p_t-1)=p_t(p_t+2p_0-1)=p_t^2-p_t+2(p_0^2-p_0)=0
\end{equation}
These equations show that $p_t(r)$ and $p_0(r)$ are both constants. There are three group solutions, which are $\{p_0=0, p_t=0\}, \{p_0=0, p_t=1\}$ and $\{p_0=2/3, p_t=-1/3\}$. Only the last one gives a solution which has singularity at $r=0$. One can check this solution satisfies the Kasner's relationships,
\begin{equation}\label{Kasner1}
  2p_0^2+p_t^2=2p_0+p_t=1.
\end{equation}


\bibliography{ref}

\begin{thebibliography}{19}%
\makeatletter
\providecommand \@ifxundefined [1]{%
 \@ifx{#1\undefined}
}%
\providecommand \@ifnum [1]{%
 \ifnum #1\expandafter \@firstoftwo
 \else \expandafter \@secondoftwo
 \fi
}%
\providecommand \@ifx [1]{%
 \ifx #1\expandafter \@firstoftwo
 \else \expandafter \@secondoftwo
 \fi
}%
\providecommand \natexlab [1]{#1}%
\providecommand \enquote  [1]{``#1''}%
\providecommand \bibnamefont  [1]{#1}%
\providecommand \bibfnamefont [1]{#1}%
\providecommand \citenamefont [1]{#1}%
\providecommand \href@noop [0]{\@secondoftwo}%
\providecommand \href [0]{\begingroup \@sanitize@url \@href}%
\providecommand \@href[1]{\@@startlink{#1}\@@href}%
\providecommand \@@href[1]{\endgroup#1\@@endlink}%
\providecommand \@sanitize@url [0]{\catcode `\\12\catcode `\$12\catcode
  `\&12\catcode `\#12\catcode `\^12\catcode `\_12\catcode `\%12\relax}%
\providecommand \@@startlink[1]{}%
\providecommand \@@endlink[0]{}%
\providecommand \url  [0]{\begingroup\@sanitize@url \@url }%
\providecommand \@url [1]{\endgroup\@href {#1}{\urlprefix }}%
\providecommand \urlprefix  [0]{URL }%
\providecommand \Eprint [0]{\href }%
\providecommand \doibase [0]{http://dx.doi.org/}%
\providecommand \selectlanguage [0]{\@gobble}%
\providecommand \bibinfo  [0]{\@secondoftwo}%
\providecommand \bibfield  [0]{\@secondoftwo}%
\providecommand \translation [1]{[#1]}%
\providecommand \BibitemOpen [0]{}%
\providecommand \bibitemStop [0]{}%
\providecommand \bibitemNoStop [0]{.\EOS\space}%
\providecommand \EOS [0]{\spacefactor3000\relax}%
\providecommand \BibitemShut  [1]{\csname bibitem#1\endcsname}%
\let\auto@bib@innerbib\@empty
\bibitem [{\citenamefont {Brown}\ \emph
  {et~al.}(2016{\natexlab{a}})\citenamefont {Brown}, \citenamefont {Roberts},
  \citenamefont {Susskind}, \citenamefont {Swingle},\ and\ \citenamefont
  {Zhao}}]{Brown:2015bva}%
  \BibitemOpen
  \bibfield  {author} {\bibinfo {author} {\bibfnamefont {A.~R.}\ \bibnamefont
  {Brown}}, \bibinfo {author} {\bibfnamefont {D.~A.}\ \bibnamefont {Roberts}},
  \bibinfo {author} {\bibfnamefont {L.}~\bibnamefont {Susskind}}, \bibinfo
  {author} {\bibfnamefont {B.}~\bibnamefont {Swingle}}, \ and\ \bibinfo
  {author} {\bibfnamefont {Y.}~\bibnamefont {Zhao}},\ }\href {\doibase
  10.1103/PhysRevLett.116.191301} {\bibfield  {journal} {\bibinfo  {journal}
  {Phys. Rev. Lett.}\ }\textbf {\bibinfo {volume} {116}},\ \bibinfo {pages}
  {191301} (\bibinfo {year} {2016}{\natexlab{a}})},\ \Eprint
  {http://arxiv.org/abs/1509.07876} {arXiv:1509.07876 [hep-th]} \BibitemShut
  {NoStop}%
\bibitem [{\citenamefont {Brown}\ \emph
  {et~al.}(2016{\natexlab{b}})\citenamefont {Brown}, \citenamefont {Roberts},
  \citenamefont {Susskind}, \citenamefont {Swingle},\ and\ \citenamefont
  {Zhao}}]{Brown:2015lvg}%
  \BibitemOpen
  \bibfield  {author} {\bibinfo {author} {\bibfnamefont {A.~R.}\ \bibnamefont
  {Brown}}, \bibinfo {author} {\bibfnamefont {D.~A.}\ \bibnamefont {Roberts}},
  \bibinfo {author} {\bibfnamefont {L.}~\bibnamefont {Susskind}}, \bibinfo
  {author} {\bibfnamefont {B.}~\bibnamefont {Swingle}}, \ and\ \bibinfo
  {author} {\bibfnamefont {Y.}~\bibnamefont {Zhao}},\ }\href {\doibase
  10.1103/PhysRevD.93.086006} {\bibfield  {journal} {\bibinfo  {journal} {Phys.
  Rev.}\ }\textbf {\bibinfo {volume} {D93}},\ \bibinfo {pages} {086006}
  (\bibinfo {year} {2016}{\natexlab{b}})},\ \Eprint
  {http://arxiv.org/abs/1512.04993} {arXiv:1512.04993 [hep-th]} \BibitemShut
  {NoStop}%
\bibitem [{\citenamefont {Maldacena}(2003)}]{Maldacena:2001kr}%
  \BibitemOpen
  \bibfield  {author} {\bibinfo {author} {\bibfnamefont {J.~M.}\ \bibnamefont
  {Maldacena}},\ }\href {\doibase 10.1088/1126-6708/2003/04/021} {\bibfield
  {journal} {\bibinfo  {journal} {JHEP}\ }\textbf {\bibinfo {volume} {04}},\
  \bibinfo {pages} {021} (\bibinfo {year} {2003})},\ \Eprint
  {http://arxiv.org/abs/hep-th/0106112} {arXiv:hep-th/0106112 [hep-th]}
  \BibitemShut {NoStop}%
\bibitem [{\citenamefont {Lehner}\ \emph {et~al.}(2016)\citenamefont {Lehner},
  \citenamefont {Myers}, \citenamefont {Poisson},\ and\ \citenamefont
  {Sorkin}}]{Lehner:2016vdi}%
  \BibitemOpen
  \bibfield  {author} {\bibinfo {author} {\bibfnamefont {L.}~\bibnamefont
  {Lehner}}, \bibinfo {author} {\bibfnamefont {R.~C.}\ \bibnamefont {Myers}},
  \bibinfo {author} {\bibfnamefont {E.}~\bibnamefont {Poisson}}, \ and\
  \bibinfo {author} {\bibfnamefont {R.~D.}\ \bibnamefont {Sorkin}},\
  }\href@noop {} {\  (\bibinfo {year} {2016})},\ \Eprint
  {http://arxiv.org/abs/1609.00207} {arXiv:1609.00207 [hep-th]} \BibitemShut
  {NoStop}%
\bibitem [{\citenamefont {Bekenstein}(1973)}]{PhysRevD.7.2333}%
  \BibitemOpen
  \bibfield  {author} {\bibinfo {author} {\bibfnamefont {J.~D.}\ \bibnamefont
  {Bekenstein}},\ }\href {\doibase 10.1103/PhysRevD.7.2333} {\bibfield
  {journal} {\bibinfo  {journal} {Phys. Rev. D}\ }\textbf {\bibinfo {volume}
  {7}},\ \bibinfo {pages} {2333} (\bibinfo {year} {1973})}\BibitemShut
  {NoStop}%
\bibitem [{\citenamefont {Margolus}\ and\ \citenamefont
  {Levitin}(1998)}]{MARGOLUS1998188}%
  \BibitemOpen
  \bibfield  {author} {\bibinfo {author} {\bibfnamefont {N.}~\bibnamefont
  {Margolus}}\ and\ \bibinfo {author} {\bibfnamefont {L.~B.}\ \bibnamefont
  {Levitin}},\ }\href {\doibase
  http://dx.doi.org/10.1016/S0167-2789(98)00054-2} {\bibfield  {journal}
  {\bibinfo  {journal} {Physica D: Nonlinear Phenomena}\ }\textbf {\bibinfo
  {volume} {120}},\ \bibinfo {pages} {188 } (\bibinfo {year}
  {1998})}\BibitemShut {NoStop}%
\bibitem [{\citenamefont {Lloyd}(2000)}]{Lloyd2000}%
  \BibitemOpen
  \bibfield  {author} {\bibinfo {author} {\bibfnamefont {S.}~\bibnamefont
  {Lloyd}},\ }\href {\doibase 10.1038/35023282} {\bibfield  {journal} {\bibinfo
   {journal} {Nature}\ }\textbf {\bibinfo {volume} {406}},\ \bibinfo {pages}
  {1047} (\bibinfo {year} {2000})}\BibitemShut {NoStop}%
\bibitem [{\citenamefont {Abbott}\ and\ \citenamefont
  {Deser}(1982)}]{Abbott:1981ff}%
  \BibitemOpen
  \bibfield  {author} {\bibinfo {author} {\bibfnamefont {L.~F.}\ \bibnamefont
  {Abbott}}\ and\ \bibinfo {author} {\bibfnamefont {S.}~\bibnamefont {Deser}},\
  }\href {\doibase 10.1016/0550-3213(82)90049-9} {\bibfield  {journal}
  {\bibinfo  {journal} {Nucl. Phys.}\ }\textbf {\bibinfo {volume} {B195}},\
  \bibinfo {pages} {76} (\bibinfo {year} {1982})}\BibitemShut {NoStop}%
\bibitem [{\citenamefont {Misner}\ and\ \citenamefont
  {Sharp}(1964)}]{PhysRev.136.B571}%
  \BibitemOpen
  \bibfield  {author} {\bibinfo {author} {\bibfnamefont {C.~W.}\ \bibnamefont
  {Misner}}\ and\ \bibinfo {author} {\bibfnamefont {D.~H.}\ \bibnamefont
  {Sharp}},\ }\href {\doibase 10.1103/PhysRev.136.B571} {\bibfield  {journal}
  {\bibinfo  {journal} {Phys. Rev.}\ }\textbf {\bibinfo {volume} {136}},\
  \bibinfo {pages} {B571} (\bibinfo {year} {1964})}\BibitemShut {NoStop}%
\bibitem [{\citenamefont {Komar}(1963)}]{PhysRev.129.1873}%
  \BibitemOpen
  \bibfield  {author} {\bibinfo {author} {\bibfnamefont {A.}~\bibnamefont
  {Komar}},\ }\href {\doibase 10.1103/PhysRev.129.1873} {\bibfield  {journal}
  {\bibinfo  {journal} {Phys. Rev.}\ }\textbf {\bibinfo {volume} {129}},\
  \bibinfo {pages} {1873} (\bibinfo {year} {1963})}\BibitemShut {NoStop}%
\bibitem [{\citenamefont {Brown}\ and\ \citenamefont
  {York}(1993)}]{Brown:1992br}%
  \BibitemOpen
  \bibfield  {author} {\bibinfo {author} {\bibfnamefont {J.~D.}\ \bibnamefont
  {Brown}}\ and\ \bibinfo {author} {\bibfnamefont {J.~W.}\ \bibnamefont {York},
  \bibfnamefont {Jr.}},\ }\href {\doibase 10.1103/PhysRevD.47.1407} {\bibfield
  {journal} {\bibinfo  {journal} {Phys. Rev.}\ }\textbf {\bibinfo {volume}
  {D47}},\ \bibinfo {pages} {1407} (\bibinfo {year} {1993})},\ \Eprint
  {http://arxiv.org/abs/gr-qc/9209012} {arXiv:gr-qc/9209012 [gr-qc]}
  \BibitemShut {NoStop}%
\bibitem [{\citenamefont {Barnich}\ and\ \citenamefont
  {Compere}(2005)}]{Barnich:2004uw}%
  \BibitemOpen
  \bibfield  {author} {\bibinfo {author} {\bibfnamefont {G.}~\bibnamefont
  {Barnich}}\ and\ \bibinfo {author} {\bibfnamefont {G.}~\bibnamefont
  {Compere}},\ }\href {\doibase 10.1103/PhysRevD.71.044016,
  10.1103/PhysRevD.73.029904, 10.1103/PhysRevD.71.029904} {\bibfield  {journal}
  {\bibinfo  {journal} {Phys. Rev.}\ }\textbf {\bibinfo {volume} {D71}},\
  \bibinfo {pages} {044016} (\bibinfo {year} {2005})},\ \bibinfo {note}
  {[Erratum: Phys. Rev.D73,029904(2006)]},\ \Eprint
  {http://arxiv.org/abs/gr-qc/0412029} {arXiv:gr-qc/0412029 [gr-qc]}
  \BibitemShut {NoStop}%
\bibitem [{\citenamefont {Kastor}\ \emph {et~al.}(2009)\citenamefont {Kastor},
  \citenamefont {Ray},\ and\ \citenamefont {Traschen}}]{Kastor:2009wy}%
  \BibitemOpen
  \bibfield  {author} {\bibinfo {author} {\bibfnamefont {D.}~\bibnamefont
  {Kastor}}, \bibinfo {author} {\bibfnamefont {S.}~\bibnamefont {Ray}}, \ and\
  \bibinfo {author} {\bibfnamefont {J.}~\bibnamefont {Traschen}},\ }\href
  {\doibase 10.1088/0264-9381/26/19/195011} {\bibfield  {journal} {\bibinfo
  {journal} {Class. Quant. Grav.}\ }\textbf {\bibinfo {volume} {26}},\ \bibinfo
  {pages} {195011} (\bibinfo {year} {2009})},\ \Eprint
  {http://arxiv.org/abs/0904.2765} {arXiv:0904.2765 [hep-th]} \BibitemShut
  {NoStop}%
\bibitem [{\citenamefont {Parattu}\ \emph {et~al.}(2016)\citenamefont
  {Parattu}, \citenamefont {Chakraborty}, \citenamefont {Majhi},\ and\
  \citenamefont {Padmanabhan}}]{Parattu:2015gga}%
  \BibitemOpen
  \bibfield  {author} {\bibinfo {author} {\bibfnamefont {K.}~\bibnamefont
  {Parattu}}, \bibinfo {author} {\bibfnamefont {S.}~\bibnamefont
  {Chakraborty}}, \bibinfo {author} {\bibfnamefont {B.~R.}\ \bibnamefont
  {Majhi}}, \ and\ \bibinfo {author} {\bibfnamefont {T.}~\bibnamefont
  {Padmanabhan}},\ }\href {\doibase 10.1007/s10714-016-2093-7} {\bibfield
  {journal} {\bibinfo  {journal} {Gen. Rel. Grav.}\ }\textbf {\bibinfo {volume}
  {48}},\ \bibinfo {pages} {94} (\bibinfo {year} {2016})},\ \Eprint
  {http://arxiv.org/abs/1501.01053} {arXiv:1501.01053 [gr-qc]} \BibitemShut
  {NoStop}%
\bibitem [{\citenamefont {Belinskii}\ \emph {et~al.}(1970)\citenamefont
  {Belinskii}, \citenamefont {Khalatnikov},\ and\ \citenamefont
  {Lifshitz}}]{Belinskii1970}%
  \BibitemOpen
  \bibfield  {author} {\bibinfo {author} {\bibfnamefont {V.}~\bibnamefont
  {Belinskii}}, \bibinfo {author} {\bibfnamefont {I.}~\bibnamefont
  {Khalatnikov}}, \ and\ \bibinfo {author} {\bibfnamefont {E.}~\bibnamefont
  {Lifshitz}},\ }\href {\doibase 10.1080/00018737000101171} {\bibfield
  {journal} {\bibinfo  {journal} {Advances in Physics}\ }\textbf {\bibinfo
  {volume} {19}},\ \bibinfo {pages} {525} (\bibinfo {year} {1970})}\BibitemShut
  {NoStop}%
\bibitem [{\citenamefont {Belinskii}\ \emph {et~al.}(1982)\citenamefont
  {Belinskii}, \citenamefont {Khalatnikov},\ and\ \citenamefont
  {Lifshitz}}]{Belinskii1982}%
  \BibitemOpen
  \bibfield  {author} {\bibinfo {author} {\bibfnamefont {V.}~\bibnamefont
  {Belinskii}}, \bibinfo {author} {\bibfnamefont {I.}~\bibnamefont
  {Khalatnikov}}, \ and\ \bibinfo {author} {\bibfnamefont {E.}~\bibnamefont
  {Lifshitz}},\ }\href {\doibase 10.1080/00018738200101428} {\bibfield
  {journal} {\bibinfo  {journal} {Advances in Physics}\ }\textbf {\bibinfo
  {volume} {31}},\ \bibinfo {pages} {639} (\bibinfo {year} {1982})}\BibitemShut
  {NoStop}%
\bibitem [{\citenamefont {Cai}\ \emph {et~al.}(2016)\citenamefont {Cai},
  \citenamefont {Ruan}, \citenamefont {Wang}, \citenamefont {Yang},\ and\
  \citenamefont {Peng}}]{Cai:2016xho}%
  \BibitemOpen
  \bibfield  {author} {\bibinfo {author} {\bibfnamefont {R.-G.}\ \bibnamefont
  {Cai}}, \bibinfo {author} {\bibfnamefont {S.-M.}\ \bibnamefont {Ruan}},
  \bibinfo {author} {\bibfnamefont {S.-J.}\ \bibnamefont {Wang}}, \bibinfo
  {author} {\bibfnamefont {R.-Q.}\ \bibnamefont {Yang}}, \ and\ \bibinfo
  {author} {\bibfnamefont {R.-H.}\ \bibnamefont {Peng}},\ }\href {\doibase
  10.1007/JHEP09(2016)161} {\bibfield  {journal} {\bibinfo  {journal} {JHEP}\
  }\textbf {\bibinfo {volume} {09}},\ \bibinfo {pages} {161} (\bibinfo {year}
  {2016})},\ \Eprint {http://arxiv.org/abs/1606.08307} {arXiv:1606.08307
  [gr-qc]} \BibitemShut {NoStop}%
\bibitem [{\citenamefont {Huang}\ \emph {et~al.}(2016)\citenamefont {Huang},
  \citenamefont {Feng},\ and\ \citenamefont {Lu}}]{Huang:2016fks}%
  \BibitemOpen
  \bibfield  {author} {\bibinfo {author} {\bibfnamefont {H.}~\bibnamefont
  {Huang}}, \bibinfo {author} {\bibfnamefont {X.-H.}\ \bibnamefont {Feng}}, \
  and\ \bibinfo {author} {\bibfnamefont {H.}~\bibnamefont {Lu}},\ }\href@noop
  {} {\  (\bibinfo {year} {2016})},\ \Eprint {http://arxiv.org/abs/1611.02321}
  {arXiv:1611.02321 [hep-th]} \BibitemShut {NoStop}%
\bibitem [{\citenamefont {Cai}\ \emph {et~al.}(2017)\citenamefont {Cai},
  \citenamefont {Sasaki},\ and\ \citenamefont {Wang}}]{Cai:2017sjv}%
  \BibitemOpen
  \bibfield  {author} {\bibinfo {author} {\bibfnamefont {R.-G.}\ \bibnamefont
  {Cai}}, \bibinfo {author} {\bibfnamefont {M.}~\bibnamefont {Sasaki}}, \ and\
  \bibinfo {author} {\bibfnamefont {S.-J.}\ \bibnamefont {Wang}},\ }\href@noop
  {} {\  (\bibinfo {year} {2017})},\ \Eprint {http://arxiv.org/abs/1702.06766}
  {arXiv:1702.06766 [gr-qc]} \BibitemShut {NoStop}%
\end{thebibliography}%

\end{document}